\documentclass[aps, prd,11pt, tightenlines, secnumarabic, superscriptaddress,showpacs, preprintnumbers, nofootinbib]{revtex4-1}
\pdfoutput=1
\usepackage{xspace}
\usepackage{pifont}
\usepackage{soul}
\usepackage{todonotes}
\usepackage{multirow,array}
\usepackage[english]{babel}
\usepackage{amsmath,amssymb,amsfonts,bm,bbm,slashed}
\usepackage{graphicx}
\usepackage[sort&compress]{natbib}
\usepackage{xcolor}
\usepackage[normalem]{ulem}
\usepackage{hyperref}
\usepackage{cleveref}
\definecolor{red}{rgb}{1.0, 0, 0}
\usepackage{enumerate}
\usepackage{epsfig, subfigure}
\usepackage{epstopdf}
\usepackage{setspace}
\usepackage{booktabs, tabularx}
\usepackage{units}
\usepackage{tikz}
\usepackage{ctable}

\usetikzlibrary{arrows,shapes}
\usetikzlibrary{trees}
\usetikzlibrary{matrix,arrows}                          
\usetikzlibrary{positioning}                            
\usetikzlibrary{calc,through}                           
\usetikzlibrary{decorations.pathreplacing}  
\usepackage{pgffor}                                                     

\usetikzlibrary{decorations.pathmorphing}       
\usetikzlibrary{decorations.markings}
\tikzset{
  vector/.style={decorate, decoration={snake,amplitude=2.0pt}, draw=blue},
      provector/.style={decorate, decoration={snake,amplitude=2.0pt}, draw=blue},
      antivector/.style={decorate, decoration={snake,amplitude=-2.0pt}, draw=black},
  fermion/.style={draw=black, postaction={decorate},
      decoration={markings,mark=at position .55 with {\arrow[draw=black]{>}}}},
  fermionbar/.style={draw=black, postaction={decorate},
      decoration={markings,mark=at position .55 with {\arrow[draw=black]{<}}}},
  fermionnoarrow/.style={draw=red},
  gluon/.style={decorate, draw=purple,
      decoration={coil,amplitude=4pt, segment length=5pt}},
  scalar/.style={dashed,draw=red, postaction={decorate},
      decoration={markings,mark=at position .55 with {\arrow[draw=red]{<}}}},
  scalarbar/.style={dashed,draw=red, postaction={decorate},
      decoration={markings,mark=at position .55 with {\arrow[draw=red]{<}}}},
  scalarnoarrow/.style={dashed,draw=red},
  electron/.style={draw=red, postaction={decorate},
      decoration={markings,mark=at position .55 with {\arrow[draw=red]{>}}}},
      bigvector/.style={decorate, decoration={snake,amplitude=4pt}, draw=blue},
}

\hypersetup{
    pdfnewwindow=true,   
    colorlinks=true,     
    linkcolor=blue,      
    citecolor=blue,      
    filecolor=blue,      
    urlcolor=blue        
}

\newcolumntype{N}{>{\centering\arraybackslash}m{4cm}}
\newcolumntype{G}{>{\bfseries\centering\arraybackslash}m{3cm+6\tabcolsep}}
\newcolumntype{M}[1]{>{\centering\arraybackslash}m{#1}}

\allowdisplaybreaks
\DeclareMathOperator{\Tr}{Tr}

\bibpunct{[}{]}{,}{n}{}{,}
\newcommand{\ev}[1]{\ensuremath{\left\langle #1 %
                     \right\rangle}} 

\newcommand{\BR}{\text{BR}}

\definecolor{seagreen}{rgb}{0.180392,0.545098,0.341176}

\newcommand{\eps}{{\ensuremath{\varepsilon}}}

\begin{document}

\title{Impeded Dark Matter}

\author{Joachim Kopp}
\email{jkopp@uni-mainz.de}
\affiliation{PRISMA Cluster of Excellence \& Mainz Institute for Theoretical Physics,
             Johannes Gutenberg University, Staudingerweg 7, 55099 Mainz, Germany}

\author{Jia Liu}
\email{liuj@uni-mainz.de}
\affiliation{PRISMA Cluster of Excellence \& Mainz Institute for Theoretical Physics,
             Johannes Gutenberg University, Staudingerweg 7, 55099 Mainz, Germany}

\author{Tracy R. Slatyer}
\email{tslatyer@mit.edu}
\affiliation{Center for Theoretical Physics, Massachusetts Institute of Technology,
             Cambridge, MA 02139, USA}

\author{Xiao-Ping Wang}
\email{xiaowang@uni-mainz.de}
\affiliation{PRISMA Cluster of Excellence \& Mainz Institute for Theoretical Physics,
             Johannes Gutenberg University, Staudingerweg 7, 55099 Mainz, Germany}

\author{Wei Xue}
\email{weixue@mit.edu}
\affiliation{Center for Theoretical Physics, Massachusetts Institute of Technology,
             Cambridge, MA 02139, USA}

\date{\today}

\preprint{\parbox{3cm}{\flushright MITP/16-092 \\ MIT-CTP-4832}}


\begin{abstract}
  We consider dark matter models in which the mass splitting between the dark
  matter particles and their annihilation products is tiny.  Compared to the
  previously proposed Forbidden Dark Matter scenario, the mass splittings we
  consider are much smaller, and are allowed to be either positive or negative.
  To emphasize this modification, we dub our scenario ``Impeded Dark Matter''.
  We demonstrate that Impeded Dark Matter can be easily realized without
  requiring tuning of model parameters.  For negative mass splitting, we
  demonstrate that the annihilation cross-section for Impeded Dark Matter
  depends linearly on the dark matter velocity or may even be kinematically
  forbidden, making this scenario almost insensitive to constraints from the
  cosmic microwave background and from observations of dwarf galaxies.
  Accordingly, it may be possible for Impeded Dark Matter to yield observable
  signals in clusters or the Galactic center, with no corresponding signal in
  dwarfs.  For positive mass splitting, we show that the annihilation
  cross-section is suppressed by the small mass splitting, which helps light
  dark matter to survive increasingly stringent constraints from indirect
  searches.  As specific realizations for Impeded Dark Matter, we introduce a
  model of vector dark matter from a hidden $SU(2)$ sector, and a composite
  dark matter scenario based on a QCD-like dark sector.
\end{abstract}

\maketitle
\tableofcontents

\section{Introduction}
\label{sec:intro}

Many theories in particle physics live through an infancy in which they are
carved out by a few pioneering masterminds, a youth characterized by wild
enthusiasm in the broader community, an adulthood in which they become part of
university curricula, and the sunset years during which lack of experimental
evidence leads to disillusionment or at least fatigue in the community.  Models
of Weakly Interacting Massive Particles (WIMPs) may be approaching this last
stage of their life cycle and may eventually fade away unless solid
experimental evidence for WIMP dark matter (DM) is discovered soon.
Nevertheless, this time has not come yet, and in fact WIMPs are experiencing an
Indian summer with fresh ideas and models sprouting from the arXiv on a regular
basis.  Promising recent developments include Secluded DM~\cite{Pospelov:2007mp,
ArkaniHamed:2008qn}, SIMP~\cite{Hochberg:2014dra, Hochberg:2014kqa,Lee:2015gsa,
Hochberg:2015vrg}, Selfish DM~\cite{D'Agnolo:2015pha}, Forbidden
DM~\cite{D'Agnolo:2015koa,Delgado:2016umt}, Cannibal DM~\cite{Carlson:1992fn,
  Pappadopulo:2016pkp, Bernal:2015ova, Kuflik:2015isi, Bernal:2015xba,
Farina:2016llk}, Co-decaying DM~\cite{Dror:2016rxc, Okawa:2016wrr, Bandyopadhyay:2011qm},
Semi-annihilating DM~\cite{D'Eramo:2010ep}, Boosted DM~\cite{Agashe:2014yua,
  Berger:2014sqa, Kopp:2015bfa},
and DM with late-time dilution \cite{Berlin:2016vnh}.
These scenarios are characterized by a dark matter sector with non-minimal
particle content and interesting, unconventional dynamics.

This is also true for the scenarios we wish to consider in the present work.
In particular, we consider situations in which the dynamics of DM in the early
Universe is governed by a dominant annihilation channel $\text{DM} \, \text{DM}
\to \text{X} \, \text{X}$, with the special feature that the mass splitting
$\Delta \equiv m_\text{DM} - m_\text{X}$
between $\text{DM}$ and $\text{X}$ is very small, $|\Delta| \ll m_\text{DM}$.
Our scenario is thus closely related to  Forbidden
DM~\cite{D'Agnolo:2015koa,Delgado:2016umt} and Co-decaying
DM~\cite{Dror:2016rxc, Okawa:2016wrr, Bandyopadhyay:2011qm}.
We allow $\Delta$ to be either positive or negative, and we assume that
$\text{X}$ couples also to SM particles. The dynamics of DM annihilation in the
non-relativistic regime is governed by the phase space factor
in the cross-section, therefore we call this scenario ``Impeded Dark
Matter''.
Explicitly, the velocity-averaged annihilation cross-section for Impeded DM
has the form (see also \cite{Jia:2016pbe})
\begin{align}
  \ev{\sigma v_\text{rel}}
    =      \int \frac{ d \sigma_0}{ d \Omega} \frac{1}{4 \pi}
           \sqrt{1 - \frac{4 m_\text{X}^2}{s}} d\Omega
    \ \simeq \ \sigma_0
           \sqrt{\frac{v_\text{rel}^2}{4} + \frac{2\Delta}{m_\text{DM}}} \,.
  \label{eq:phasespace}
\end{align}
We consider only scenarios in which $\sigma_0$ is independent of $v_\text{rel}$
at leading order, i.e.\ in which DM annihilation is an $s$-wave process.
Around the time of DM freeze-out, when $\ev{v_\text{rel}^2} \sim 0.26$
is large enough to neglect the mass difference $\Delta$, but low
enough to treat DM and $\text{X}$ as non-relativistic, we obtain
$\ev{\sigma v_\text{rel}} \simeq \sigma_0 v_\text{rel}/2$.
This linear dependence on $v_\text{rel}$ distinguishes Impeded DM from
most other DM models, in which $\ev{\sigma v_\text{rel}}$ is either
velocity-independent or proportional to $v_\text{rel}^2$. (Scenarios where
phase space suppression leads to a linear dependence of $\ev{\sigma v_\text{rel}}$
on $v_\text{rel}$ have also been discussed recently for instance in
refs.~\cite{Dror:2016rxc, Jia:2016pbe, Okawa:2016wrr}.)

\begin{figure*}
  \begin{tabular}{c@{\hspace{-1cm}}c}
    \bf{SU(2)} &\bf{Composite} \\
    \hspace{2cm}\includegraphics[width=0.4\columnwidth]{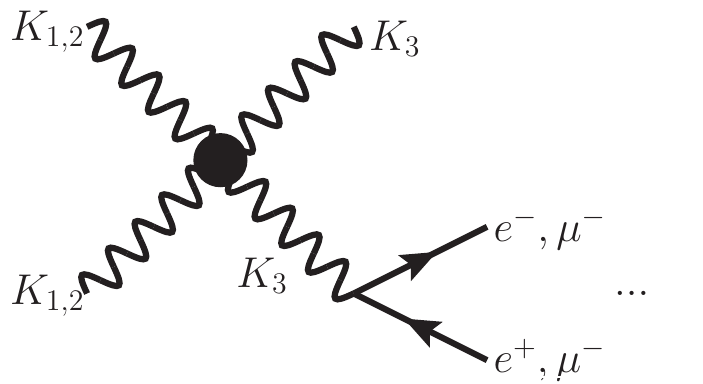} &
    \hspace{2cm}\includegraphics[width=0.4\columnwidth]{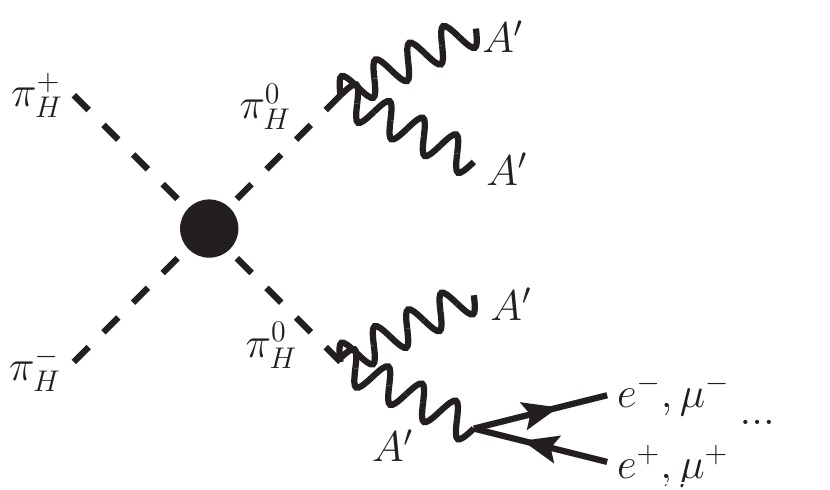} \\
    (a) & (b)
  \end{tabular}
  \caption{Dark matter annihilation in two Impeded DM models: (a)
    a dark $SU(2)$ model with $\Delta \equiv m_\text{DM} - m_\text{X} < 0$,
    where $K_{1,2,3}$ are the gauge
    bosons associated with the new gauge symmetry. $K_1$ and $K_2$ are
    degenerate in mass, while $K_3$ is slightly heavier thanks to a
    higher-dimensional coupling.  The same coupling also mixes $K_3$ with the
    SM hypercharge boson $B$. (b) a QCD+QED-like composite dark sector with
    $\Delta > 0$, in which the mass-degenerate charged dark pions $\pi^\pm_H$
    act as the dark matter, while their neutral partner $\pi^0_H$ can decay to
    two dark photons $A'$ through the chiral anomaly.}
  \label{fig:freezeout2}
\end{figure*}

From a theorist's point of view, the small mass splitting $\Delta$ can be
explained easily, for instance if the DM and $\text{X}$ are members of the same
multiplet under a gauge symmetry, global symmetry, or supersymmetry (SUSY).
Once the symmetry is broken by a small amount---as is desirable to allow
$\text{X}$ to decay to SM particles---$m_\text{DM}$ and $m_\text{X}$ become
split, either at tree level or through loop effects.  In the following, we
investigate in particular small mass splittings arising from a dark sector
gauge symmetry $SU(2)_d$, or from an approximate global chiral symmetry in a
composite hidden sector.  The annihilation processes for both cases are
illustrated in \cref{fig:freezeout2}.  We will not discuss SUSY here, but we
remark that Impeded DM can be easily realized in stealth
SUSY~\cite{Fan:2011yu,Fan:2012jf,Fan:2015mxp} under the condition that $\Delta$
is larger than the gravitino mass.

We summarize the main phenomenological features that distinguish Impeded DM models
from other scenarios.
\begin{itemize}
  \item The linear dependence of the annihilation cross-section on the DM
    velocity is crucial for indirect detection: it leads to strong signals in
    regions of larger DM velocity such as galaxy clusters or the Galactic center,
    while signals from objects with low DM velocity dispersion, in particular
    dwarf galaxies, are suppressed.

  \item If $\Delta < 0$ (DM lighter than $\text{X}$), annihilation to $XX$ becomes
    kinematically forbidden at too low DM velocity.  This is phenomenologically
    relevant when $\ev{v_\text{rel}^2} < 8|\Delta|/m_\text{DM}$, in which case
    the cross-section $\ev{\sigma v_\text{rel}}$ is Boltzmann suppressed.
    Typical values for $\ev{v_\text{rel}^2}$ are $10^{-9}$ in dwarf galaxies,
    $10^{-6}$ in the Milky Way, $10^{-5}$ in galaxy clusters, $0.26$ at freeze-out,
    and $< 10^{-9}\,\text{GeV}/m_\text{DM}$ at the epoch of last scattering
    relevant to the CMB limits.\footnote{If kinetic
    decoupling between the dark and visible sectors occurs before the
    epoch of last scattering, as is usually the case, the DM temperature will be even
    smaller. The reason is that, after kinetic decoupling, the dark sector
    temperatures evolves as $a^{-2}(t)$, which the photon temperature
    drops only as $a^{-1}(t)$, where $a(t)$ is the scale factor of the Universe.}

  \item For $\Delta > 0$, the parametric dependence of the
    annihilation cross-section changes at very low velocity, making $\ev{\sigma
    v_\text{rel}}$ dominated by the mass splitting and independent of
    $v_\text{rel}$. DM annihilation is never kinematically forbidden in this
    case.
\end{itemize}

The remainder of this paper is organized as follows: In \cref{sec:darkSU2}, we
introduce a dark $SU(2)$ model as a promising example for Impeded DM with
$\Delta < 0$. The setup of this model is discussed in \cref{sec:SU2model}.  We
calculate the relic abundance in \cref{sec:relic-density}, and discuss the
direct detection, CMB and indirect constraints in \cref{sec:dd}, \cref{sec:cmb}
and \cref{sec:indirectD} respectively.  As an example for $\Delta > 0$, we then
study a dark pion model in \cref{sec:dark-pion}. Once again, we commence by
introducing the model in \cref{sec:compositemodel}, and then investigate its
freeze-out dynamics and detection prospects in \cref{sec:darkpion-constraints}.
In \cref{sec:conclusions}, we conclude.

\section{Dark $\text{SU}(2)$ Gauge Bosons as Impeded DM with $\Delta < 0$}
\label{sec:darkSU2}

\subsection{Model}
\label{sec:SU2model}

Impeded Dark Matter is realized most easily when the DM particle and its
annihilation partner are members of the same multiplet under a symmetry group,
so that, for unbroken symmetry, their masses are exactly equal.  Let us
consider in particular a dark sector governed by a dark $SU(2)_d$ symmetry,
with the associated gauge bosons accounting for the DM and its annihilation
products. $SU(2)_d$ is broken by a scalar doublet $\Phi = (G^1 + i G^2, (v_d +
\phi)/\sqrt{2} + i G^3)$, where, $G^1$, $G^2$, $G^3$ are Goldstone bosons, $v_d
= 2 \sqrt{\ev{\Phi^\dag \Phi}}$ is the vacuum expectation value (vev), and
$\phi$ is a physical dark Higgs boson.  The dark sector Lagrangian is then,
\begin{align}
  \mathcal{L} &= - \frac{1}{4} K_{\mu\nu}^a K_{\mu\nu}^a
                 + (D_\mu \Phi)^\dag  (D_\mu \Phi)
                 - V(\Phi) \,,
  \label{eq:L-dark}
\end{align}
with the potential
\begin{align}
  V(\Phi) &\equiv -\mu^2 \Phi^\dag \Phi + \frac{\lambda}{2} (\Phi^\dag \Phi)^2
  \,
\end{align}
and the field strength tensor $K_{\mu\nu}^a = \partial_\mu K_\nu^a -
\partial_\nu K_\mu^a + g_d \varepsilon^{abc} K_\mu^b K_\nu^c$, where $g_d$
is the $SU(2)_d$ coupling constant.
The three $SU(2)_d$ gauge fields $K_\mu^1$, $K_\mu^2$ and $K_\mu^3$
initially obtain equal masses
\begin{align}
  m_k = \frac{g_d v_d}{2}
  \label{eq:mk}
\end{align}
due to a residual global $SO(3)$ symmetry.

A dark sector with gauge boson DM can couple to the SM in various different
ways.  In most models considered in the literature, the dark and visible
sectors are connected through Higgs portal interactions~\cite{Hambye:2008bq,
  Hambye:2009fg,Farzan:2012hh,Baek:2012se, Baek:2013dwa, Ko:2014gha,
Baek:2014goa, Baek:2014jga, Chen:2015nea, Gross:2015cwa, Kim:2015hda,
DiChiara:2015bua, Chen:2015dea, Kim:2015hda, Karam:2016rsz}. Some models
instead feature particles from the dark or visible sector that are charged
under both SM and hidden gauge symmetries \cite{DiazCruz:2010dc,
Bhattacharya:2011tr, Chiang:2013kqa, Fraser:2014yga}, use Abelian kinetic
mixing between a hidden $U(1)'$ gauge boson and the SM hypercharge boson
$B^\mu$ \cite{Davoudiasl:2013jma}, or invoke loop processes and higher
dimensional operators \cite{DiFranzo:2015nli, Dror:2016rxc} to connect the two
sectors.  In order to avoid introducing extra particles, we will here consider
only the renormalizable Higgs portal interaction
\begin{align}
  \mathcal{L}_\text{Higgs portal} \equiv \lambda_p (\Phi^\dag \Phi) (H^\dag H) \,,
  \label{eq:L-H-portal}
\end{align}
and non-Abelian kinetic mixing of the form
\begin{align}
  \mathcal{L}_\text{mix}
    = \frac{1}{\Lambda^2} (\Phi^\dag T^a \Phi) K_{\mu\nu}^a B_{\mu\nu}
  \label{eq:nonAkm}
\end{align}
at the non-renormalizable level.\footnote{Other operators like
$\frac{1}{\Lambda^2} (\Phi^\dag D^\mu \Phi)(H^\dag D_\mu H)$ can also
contribute to mixing \cite{Dror:2016rxc}, but
if the heavy fermions generating \cref{eq:nonAkm} do not carry $SU(2)_L$ quantum numbers,
these operators will not be generated.}
Here, $B_{\mu\nu}$ is the field strength
tensor of SM hypercharge. Non-Abelian kinetic mixing allows $K_3$ to couple
to, and decay into, SM particles. The operator in \cref{eq:nonAkm} could
arise for instance from a box loop involving heavy vector-like
$SU(2)_d$ doublet fermions charged under SM hypercharge and another
$SU(2)_d$ singlet heavy fermion carrying the same SM hypercharge.

We will assume kinetic mixing between $\phi$ and the SM Higgs boson $h$
(\cref{eq:L-H-portal}) to be
small compared to the mixing between gauge bosons from \cref{eq:nonAkm}.
As long as $m_\phi > 2 m_k$, an assumption we will make in the following,
\cref{eq:L-H-portal} is not needed to allow $\phi$ to decay. Instead, the
dominant $\phi$ decay will be $\phi \to K_i K_i$ ($i=1,2,3$).

After $SU(2)_d$ breaking, the non-Abelian kinetic mixing term takes the form
\begin{align}
  \mathcal{L}_\text{mix} &\supset
    \frac{\varepsilon}{2} \bigg(1 + \frac{\phi}{{v_d}} \bigg)^2
    \big[ \partial_\mu K_\nu^3  -  \partial_\nu K_\mu^3
        + g_d (K_\mu^1 K_\nu^2 - K_\mu^2 K_\nu^1) \big] \frac{1}{\cos \theta_{w}}B_{\mu \nu} \,,
  \label{eq:L-mix}
\end{align}
with $\eps \equiv -v_d^2  \cos \theta_{w} / (2\Lambda^2)$, where $\theta_{w}$ is the Weinberg angle.
We see that mixing affects the kinetic terms of $K_\mu^3$ and
$B_\mu$, while $K_\mu^{1,2}$ are unaffected.
To move to the physical field basis, we redefine
\begin{align}
  \begin{split}
    K_\mu^3 &\to  \frac{1}{ \sqrt{1 - \frac{1}{4} \frac{\varepsilon^2}{ \cos^2 \theta_{w}}}} K_\mu^3 \\
    B_\mu   &\to B_\mu - \frac{\varepsilon}{ \cos \theta_{w}}  \frac{1}{2 \sqrt{1
    - \frac{1}{4} \frac{\varepsilon^2}{ \cos^2 \theta_{w}}}}  K_\mu^3
  \end{split} \,,
  \label{eq:su2-shifts}
\end{align}
thus removing kinetic mixing and properly normalizing the kinetic terms. We then
apply a unitary transformation to diagonalize the gauge boson mass matrix.
Henceforth, we will use the notation $K^3_\mu$, $Z_\mu$ and $A_\mu$ to
refer to the \emph{physical} neutral gauge bosons.
The mass of the physical $K^3_\mu$ is shifted by a term
proportional to $\varepsilon^2$ relative to \cref{eq:mk}:
\begin{align}
  m_{K_3}^2 &= (m_{k} - \Delta )^2
             = m_k^2
               \bigg(1 +
                 \frac{\varepsilon^2}{ \cos^2 \theta_{w}}
                 \frac{ (m_k^2 - \cos^2 \theta_{w} m_{Z,\text{SM}}^2)}
                      {m_k^2 - m_{Z,\text{SM}}^2}
               \bigg) \,,
\end{align}
and thus
\begin{align}
  \Delta \equiv m_k - m_{K_3} \simeq -\frac{m_k}{2}
  \frac{\varepsilon^2}{ \cos^2 \theta_{w}}
    \frac{ (m_k^2 - \cos^2 \theta_{w} m_{Z,\text{SM}}^2)}
         {m_k^2 - m_{Z,\text{SM}}^2}
  \label{eq:su2-delta}
\end{align}
In this expression, $\theta_{w}$ is the weak mixing angle in the
$\varepsilon \to 0$ limit and $m_{Z,\text{SM}}$ is the $Z$ boson mass in that limit.
We see that $\Delta > 0$ is possible only in a narrow mass window in which
$ m_{W,\text{SM}} < m_k < m_{Z,\text{SM}}$.
For
$\varepsilon \neq 0$, the $Z$ boson mass is shifted to $m_Z^2 = m_{Z,\text{SM}}^2 \big[
  1 + (\varepsilon^2 \tan^2 \theta_{w} m_{Z,\text{SM}}^2) / (m_{Z,\text{SM}}^2
- m_k^2) \big]$.  The coupling of $K^3_\mu$ to the SM electromagnetic and neutral weak
currents $J_\text{em}^\mu$ and $J_Z$ is given by
\begin{align}
  \mathcal{L} \supset K_3^\mu  \bigg(
                        \varepsilon \, e J_\text{em}^\mu
                      - \varepsilon g \tan\theta_{w} \frac{m_k^2}{m_k^2 - m_Z^2}
                             J_Z^\mu  \bigg) \,.
  \label{eq:k3-smcurrent}
\end{align}
Note that \cref{eq:L-mix} implies a derivative coupling between $K_3$, $\phi$,
and the photon, as well as couplings of $K_\mu^1$, $K_\mu^2$ to the photon and
the $Z$. The $K_1 K_2 \gamma$ coupling can be interpreted as a DM magnetic dipole
moment.  These operators lead to the annihilation processes $K_{1} K_{1}, K_{2}
K_{2}\to K_3 \gamma$, $K_1 K_2 \to \phi\gamma$, and $K_{1} K_{1}, K_{2} K_{2}
\to \gamma\gamma$, which are phenomenologically interesting as they feature
mono-energetic photons (see also ref.~\cite{D'Eramo:2012rr}).

\subsection{Relic Density}
\label{sec:relic-density}

\begin{table}
  \centering
  \footnotesize
  \centering
\begin{tabular}{m{6.0cm}<{\centering}m{1.8cm}<{\centering}m{1.7cm}<{\centering}m{2.0cm}<{\centering}m{2.0cm}<{\centering}m{2.0cm}<{\centering} }
\hline \hline
process &  $v_\text{rel}$-dependence & $\eps$-dependence&  freeze-out & CMB & Indirect Detection \\
\hline
\begin{tikzpicture}[line width=1.0 pt, scale=1.0]
  \draw[vector] (0,0.65)--(0.45,0);
  \draw[vector] (0,-0.65)--(0.45,0);
  \draw[scalarnoarrow] (0.45,0)--(0.9,0);
  \draw[vector] (0.9,0)--(1.35,0.65);
  \draw[vector] (0.9,0)--(1.35,-0.65);
  \node at (-0.2,0.65) {$K_1$};
  \node at (-0.2,-0.65) {$K_1$};
  \node at (0.65,0.2) {$\phi$};
  \node at (1.55,0.65) {$K_3$};
  \node at (1.55,-0.65) {$K_3$};

  \draw[vector] (2.3,0.65)--(2.75,0.25);
  \draw[vector] (3.2,0.75)--(2.75,0.25);
  \draw[vector] (2.75,0.25)--(2.75,-0.25);
  \draw[vector] (2.75,-0.25)--(2.3,-0.65);
  \draw[vector] (2.75,-0.25)--(3.2,-0.65);
  \node at (2.1,0.65) {$K_1$};
  \node at (3.4,0.65) {$K_3$};
  \node at (2.45,0.0) {$K_2$};
  \node at (2.1,-0.65) {$K_1$};
  \node at (3.4,-0.65) {$K_3$};

  \draw[vector] (4.15,0.65)--(4.6,0.);
  \draw[vector] (4.15,-0.65)--(4.6,0.);
  \draw[vector] (5.0,0.65)--(4.6,0);
  \draw[vector] (5.0,-0.65)--(4.6,0);
  \node at (3.95,0.65) {$K_1$};
  \node at (3.95,-0.65) {$K_1$};
  \node at (5.25,0.65) {$K_3$};
  \node at (5.25,-0.65) {$K_3$};
\end{tikzpicture}
        & $ \sqrt{ \frac{v_\text{rel}^2}{4} + \frac{2\Delta }{m_\text{DM}} } $
        & $1$
        & dominant
        & negligible
        & \ding{51}
\\ \hline
\begin{tikzpicture}[line width=1.0 pt, scale=1.]
  \draw[vector](0,0.65)--(0.45,0);
  \draw[vector](0,-0.65)--(0.45,0);
  \draw[scalarnoarrow] (0.45,0)--(0.9,0);
  \draw[vector] (0.9,0)--(1.35,0.65);
  \draw[antivector] (0.9,0)--(1.35,-0.65);
  \node at (-0.2,0.65) {$K_1$};
  \node at (-0.2,-0.65) {$K_1$};
  \node at (0.65,0.2) {$\phi$};
  \node at (1.55,0.65) {$K_3$};
  \node at (1.55,-0.65) {$\gamma$};

  \draw[vector] (2.3,0.65)--(2.75,0.25);
  \draw[vector] (3.2,0.65)--(2.75,0.25);
  \draw[vector] (2.75,0.25)--(2.75,-0.25);
  \draw[vector] (2.75,-0.25)--(2.3,-0.65);
  \draw[antivector] (2.75,-0.25)--(3.2,-0.65);
  \node at (2.1,0.65) {$K_1$};
  \node at (3.4,0.65) {$K_3$};
  \node at (2.45,0.0) {$K_2$};
  \node at (2.1,-0.65) {$K_1$};
  \node at (3.4,-0.65) {$\gamma$};

 \draw[vector] (4.15,0.65)--(4.6,0.25);
  \draw[antivector] (5.05,0.65)--(4.6,0.25);
  \draw[vector] (4.6,0.25)--(4.6,-0.25);
  \draw[vector] (4.15,-0.65)--(4.6,-0.25);
  \draw[vector] (5.05,-0.65)--(4.6,-0.25);
  \node at (3.95,0.65) {$K_1$};
  \node at (3.95,-0.65) {$K_1$};
   \node at (4.3,0.0) {$K_2$};
  \node at (5.25,0.65) {$\gamma$};
  \node at (5.25,-0.65) {$K_3$};
\end{tikzpicture}
        & 1
        & $\eps^2$
        & subdominant
        & dominant
        & \ding{51}\newline\small ($\gamma$ line)
\\ \hline
\begin{tikzpicture}[line width=1.0 pt, scale=1.]
  \draw[vector] (0,0.65)--(0.45,0);
  \draw[vector] (0,-0.65)--(0.45,0);
  \draw[vector] (0.45,0)--(0.9,0);
  \draw[scalarnoarrow] (0.9,0)--(1.35,0.65);
  \draw[antivector] (0.9,0)--(1.35,-0.65);
  \node at (-0.2,0.65) {$K_1$};
  \node at (-0.2,-0.65) {$K_2$};
  \node at (0.65,0.2) {$K_3$};
  \node at (1.55,0.65) {$\phi$};
  \node at (1.55,-0.65) {$\gamma$};

  \draw[vector] (2.4,0.65)--(2.85,0.25);
  \draw[scalarnoarrow] (3.3,0.65)--(2.85,0.25);
  \draw[vector] (2.85,0.25)--(2.85,-0.25);
  \draw[vector] (2.85,-0.25)--(2.4,-0.65);
  \draw[antivector] (2.85,-0.25)--(3.3,-0.65);
  \node at (2.05,0.65) {$K_{1/2}$};
  \node at (3.5,0.65) {$\phi$};
  \node at (2.4,0.0) {$K_{1/2}$};
  \node at (2.05,-0.65) {$K_{2/1}$};
  \node at (3.5,-0.65) {$\gamma$};

  \draw[vector] (4.15,0.65)--(4.6,0.);
  \draw[vector] (4.15,-0.65)--(4.6,0.);
  \draw[scalarnoarrow] (5.0,0.65)--(4.6,0);
  \draw[antivector] (5.0,-0.65)--(4.6,0);
  \node at (3.95,0.65) {$K_1$};
  \node at (3.95,-0.65) {$K_2$};
  \node at (5.25,0.65) {$\phi$};
  \node at (5.25,-0.65) {$\gamma$};
\end{tikzpicture}
        & 1
        & $\eps^2$
        & subdominant\newline\small (requires $m_\phi < 2 m_k$)
        & dominant\newline\small (requires $m_\phi < 2 m_k$)
        & \ding{51}\newline\small ($\gamma$ line if $m_\phi < 2 m_k$)
\\ \hline
\begin{tikzpicture}[line width=1.0 pt, scale=1.]

  \draw[vector] (2.3,0.65)--(2.75,0.25);
  \draw[antivector] (3.2,0.75)--(2.75,0.25);
  \draw[vector] (2.75,0.25)--(2.75,-0.25);
  \draw[vector] (2.75,-0.25)--(2.3,-0.65);
  \draw[antivector] (2.75,-0.25)--(3.2,-0.65);
  \node at (2.1,0.65) {$K_1$};
  \node at (3.4,0.65) {$\gamma$};
  \node at (2.45,0.0) {$K_2$};
  \node at (2.1,-0.65) {$K_1$};
  \node at (3.4,-0.65) {$\gamma$};

\end{tikzpicture}
        & 1
        & $\eps^4$
        & negligible
        & negligible
        & negligible
\\ \hline
\begin{tikzpicture}[line width=1.0 pt, scale=1.]
  \draw[vector] (0,0.65)--(0.45,0);
  \draw[vector] (0,-0.65)--(0.45,0);
  \draw[vector] (0.45,0)--(0.9,0);
  \draw[antivector] (0.9,0)--(1.35,0.65);
  \draw[antivector] (0.9,0)--(1.35,-0.65);
  \node at (-0.2,0.65) {$K_1$};
  \node at (-0.2,-0.65) {$K_2$};
  \node at (0.65,0.2) {$K_3$};
  \node at (1.55,0.75) {$W^+$};
  \node at (1.55,-0.75) {$W^-$};

  \draw[vector] (2.3,0.65)--(2.75,0);
  \draw[vector] (2.3,-0.65)--(2.75,0);
  \draw[antivector] (2.75,0)--(3.25,0);
  \draw[antivector] (3.25,0)--(3.7,0.65);
  \draw[antivector] (3.25,0)--(3.7,-0.65);
  \node at (2.1,0.65) {$K_1$};
  \node at (2.1,-0.65) {$K_2$};
  \node at (3.0,0.2) {$\gamma/Z$};
  \node at (4.0,0.65) {$W^+$};
  \node at (4.0,-0.65) {$W^-$};
\end{tikzpicture}
        & $v_\text{rel}^2$
        & $\eps^2$
        & subdominant
        & negligible
        & negligible
\\ \hline
\begin{tikzpicture}[line width=1.0 pt, scale=1.]
  \draw[vector] (0,0.65)--(0.45,0);
  \draw[vector] (0,-0.65)--(0.45,0);
  \draw[vector] (0.45,0)--(0.9,0);
  \draw[fermion] (0.9,0)--(1.35,0.65);
  \draw[fermionbar] (0.9,0)--(1.35,-0.65);
  \node at (-0.2,0.65) {$K_1$};
  \node at (-0.2,-0.65) {$K_2$};
  \node at (0.65,0.2) {$K_3$};
  \node at (1.55,0.65) {$f$};
  \node at (1.55,-0.65) {$\bar{f}$};

 \draw[vector] (2.3,0.65)--(2.75,0);
  \draw[vector] (2.3,-0.65)--(2.75,0);
  \draw[antivector] (2.75,0)--(3.25,0);
  \draw[fermion] (3.25,0)--(3.7,0.65);
  \draw[fermionbar] (3.25,0)--(3.7,-0.65);
  \node at (2.1,0.65) {$K_1$};
  \node at (2.1,-0.65) {$K_2$};
  \node at (3.0,0.2) {$\gamma/Z$};
  \node at (4.0,0.65) {$f$};
  \node at (4.0,-0.65) {$\bar{f}$};
\end{tikzpicture}
        & $v_\text{rel}^2$
        & $\eps^2$
        & subdominant
        & negligible
        & negligible
         \\
\hline \hline
\end{tabular}
  \caption{The dominant DM annihilation processes of the DM particles $K_{1,2}$
    in the $SU(2)_d$ model. Note that the channel
    $K_1 K_2 \to \phi \gamma$ is kinematically not accessible for $m_\phi \gtrsim 2 m_k$.
    We list (from left to right), the Feynman diagrams contributing to a given
    process, its dependence on the relative velocity $v_\text{rel}$ of the
    annihilating DM particles, its possible suppression by powers of the kinetic mixing
    parameter $\eps$, and its relevance for DM freeze-out, CMB constraints,
    indirect and direct detection.}
  \label{tab:classification}
\end{table}

In the following, we investigate the DM relic density, in the
$SU(2)_d$ model introduced in \cref{sec:SU2model} as a function of the model
parameters.  To do so, we need to solve the Boltzmann equations describing
$K_{1,2}$ annihilation and $K_3$ decay in the early Universe.  The most
relevant DM annihilation process is $K_{1} K_{1}, K_{2} K_{2} \to K_3 K_3$, the main
properties of which are summarized in the first row of
\cref{tab:classification}.  Other annihilation channels, in particular
$K_{1} K_{1}, K_{2} K_{2} \to K_3 \gamma$, $K_1 K_2 \to \phi\gamma$, and $K_1 K_2 \to f\bar{f},
W^+ W^-$ (see \cref{tab:classification}) are all suppressed by $\eps^2$.
We do not consider $K_{1} K_{1}, K_{2} K_{2} \to \gamma\gamma $ in the
calculation because it is suppressed by a factor $\eps^4$, but we still list it in
\cref{tab:classification}.  We will also disregard
$K_1 K_2 \to \phi\gamma$ in the following, assuming $m_\phi >
2 m_k$. Finally, we neglect three-body annihilation processes like
$K_1 K_1 \to K_3 K^*_3 \to K_3 \bar f f$. Their cross-sectionss are
suppressed by $\eps^2$ and by three-body phase space, and are therefore expected to
be even smaller than those for annihilation to monoenergetic photon,
$K_3 \gamma$ and $\phi\gamma$.

The $K_3$ particles produced in $K_{1,2}$ annihilation decay to SM particles
through their kinetic mixing. We will assume that this decay is faster than the
Hubble rate at $T \lesssim m_k$, which is the case if
\begin{equation}
  \eps \gtrsim \sqrt{\frac{\, g_{*}^{1/2}}{\alpha_\text{em}}
               \frac{m_k}{M_\text{Pl}}} \,
  \label{eq:Gamma-K3-condition}
\end{equation}
where $g_*$ is the total number of relativistic degrees of freedom in the
Universe, $\alpha_\text{em}$ is the electromagnetic fine structure constant,
and $M_\text{Pl}$ is the Planck mass.  If \cref{eq:Gamma-K3-condition} is
fulfilled, the number density of $K_3$ follows its equilibrium value most of
the time during DM freeze-out, but will deviate from equilibrium when $n_{K_3}$
is very small and its decay is balanced by residual DM annihilation.  If this
is not the case, freeze-out can be significantly delayed and requires a
significant increase in annihilation cross-section~\cite{Dror:2016rxc,
Okawa:2016wrr}.  $3 \to 2$ or $4 \to 2$ processes can also play an important
role in reducing the DM abundance if the coupling $g_d$ is
large~\cite{Dror:2016rxc}.  In this regime, where \cref{eq:Gamma-K3-condition}
is violated, the temperature of the dark sector deviates significantly from
that of the SM sector and evolves as $\sim \log a$, where $a$ is the scale
factor of the Universe. In our calculation, we focus on the parameter region
where \cref{eq:Gamma-K3-condition} is fulfilled.

The decoupling of DM from the thermal bath is described by the following
coupled Boltzmann equations:
\begin{align}
  \dot{n}_{12} + 3 H n_{12}
    &= -\frac{1}{2} \ev{\sigma v}_{11 \to 33}
        \bigg[ n_{12}^2 - n_3^2 \bigg( \frac{n_{12}^\text{eq}}{n_3^\text{eq}} \bigg)^2 \bigg]
      - \frac{1}{2} \ev{\sigma v}_{11 \to 3\gamma}
        \bigg[ n_{12}^2 - (n_{12}^\text{eq})^2 \frac{n_{3}}{n_3^\text{eq}} \bigg]
                                                    \nonumber \\
    &\quad
       -\frac{1}{2} \ev{\sigma v}_{12 \to f \bar{f}, W^+ W^-}
       \big[ n_{12}^2 - (n_{12}^\text{eq})^2 \big] \,,
    \label{eq:boltzmann1}\\
  \dot{n}_3 + 3 H n_3
    &= \frac{1}{2} \ev{\sigma v}_{11 \to 33}
       \bigg[ n_{12}^2 - n_3^2 \bigg( \frac{n_{12}^\text{eq}}{n_3^\text{eq}} \bigg)^2 \bigg]
     + \frac{1}{4} \ev{\sigma v}_{11 \to 3\gamma}
       \bigg[ n_{12}^2 - (n_{12}^\text{eq})^2 \frac{n_{3}}{n_3^\text{eq}} \bigg]
     - \Gamma_{K_3} \, \big[n_3 - n_3^\text{eq} \big]
  \label{eq:boltzmann2}
\end{align}
where $n_{12}$ is the total number density of DM particles ($K_1$ and $K_2$
combined), $n_3$ is the number density of $K_3$, $\Gamma_{K_3}$ is the $K_3$ decay
rate, and the thermally averaged annihilation cross-sections $\ev{\sigma
v}_{11 \to 33}$, $\ev{\sigma v}_{11 \to 3\gamma}$, and $\ev{\sigma v}_{12 \to f \bar{f}, W^+ W^-}$
correspond to the processes $K_{1} K_{1} \to K_3 K_3$,
$K_{1} K_{1} \to
K_3 \gamma$, and $K_1 K_2 \to f\bar{f}, W^+W^-$, respectively.
The annihilation cross-sections for $K_2 K_2 \to K_3 K_3$,
and $K_2 K_2 \to K_3 \gamma$ are identical to the ones for
$K_1 K_1 \to K_3 K_3$ and
$K_1 K_1 \to K_3 \gamma$, respectively.
Explicitly, we have
\begin{align}
  (\sigma v_\text{rel})_{11 \to 33}
    &= \frac{g_d^4}{3072 \pi m_k^2}
       \frac{2723 - 5472  x_\phi + 2752 x_\phi^2}{(1 - x_\phi)^2}
       \sqrt{\frac{v_\text{rel}^2}{4}
           + \frac{m_k^2 - m_{K_3}^2}{m_k^2} } \,,
  \label{eq:1133}
\\
  (\sigma v_\text{rel})_{11\to 3\gamma}
    &= \frac{9 g_d^4 \eps^2} {1024 \pi m_k^2}
       \frac{31 - 68 x_\phi + 38 x_\phi^2}{(1 - x_\phi)^2} \,,
  \label{eq:113gamma}
\\
(\sigma v_\text{rel})_{12 \to \phi \gamma } & = \frac{{g_d^4 \varepsilon ^2 }}{{72\pi m_k^2 }}\left( {1 - x_\phi} \right)\left( {1 + x_\phi} \right)^2  \,,
\end{align}
where we have introduced the notation $x_\phi \equiv m_\phi^2 / (4 m_k^2)$.
The annihilation cross-sections for the processes $K_1 K_2 \to f\bar{f}$
and $K_1 K_2 \to W^+ W^-$ are listed in \cref{sec:appendixSM}.
Note that we include only final state species lighter than $m_k$.
The thermally averaged cross-sections $\ev{\sigma v_\text{rel}}$ are
obtained from these expressions along the lines of \cite{Gondolo:1990dk}.
The decay rate $\Gamma_{K_3}$ receives contributions from
$K_3 \to f\bar{f}$ and from $K_3 \to W^+ W^-$. The corresponding expressions
are listed in \cref{sec:appendixK3}.

The dark sector and SM sector are kept in equilibrium dominantly
through $K_3 \to f \bar{f}$ decay and its inverse. Other 2-to-2 scattering
or annihilation processes like $K_3 f \leftrightarrow \gamma f$,
$K_3 \gamma \leftrightarrow \bar f f$, $K_1 K_2 \to \bar f f$ and
$K_1 f \to K_2 f$ have a lower rate because the corresponding amplitudes
are suppressed by an extra coupling constant.
Moreover, scattering rates are typically proportional to $T \sim
m_k/20$ at freeze-out, while decay rates are $\propto m_k$.

\begin{figure*}
  \includegraphics[width=0.6\textwidth]{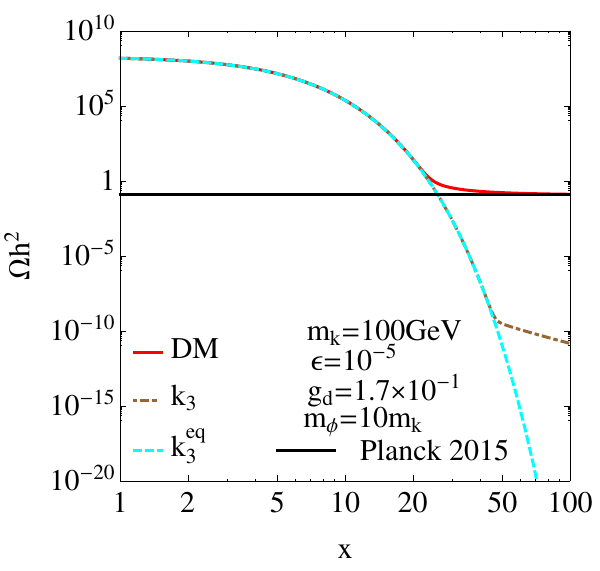}
  \caption{The evolution of the effective present day energy density $\Omega h^2$
    of DM particles
    $K_{1,2}$ and of the DM annihilation product $K_3$ for a particular choice
    of parameters in the $SU(2)_d$ model.  $\Omega h^2$ is obtained by scaling the
    instantaneous number density by the subsequent expansion of the Universe and
    normalizing to the critical density today.
    We see that the density of $K_1$, $K_2$
    (red) begins to deviate from its equilibrium value (turquoise) around $x \equiv
    m_k / T \sim 20$, while $K_3$ (brown) stays in equilibrium until
    $x \sim 50$.
    Note that for much smaller kinetic mixing $\eps$, the $K_3$ decay rate
    can become lower than the Hubble rate, substantially delaying DM freeze-out.
    (This situation was dubbed ``co-decaying DM'' in ref.~\cite{Dror:2016rxc}).
  }
  \label{fig:n1n3-evolution}
\end{figure*}

In \cref{fig:n1n3-evolution}, we plot the solution to the Boltzmann equations,
 \cref{eq:boltzmann1,eq:boltzmann2}, for a specific set of model parameters as
indicated in the plot. We see that DM freezes out at around $x_f \equiv m_k /
T_f \sim 20$, similar to a conventional Weakly Interacting Massive Particle
(WIMP). (This is only true because the $K_3$ decay rate is faster than the
Hubble rate.) At late times, around $x \sim 50$, the number density $n_3$ of
$K_3$ begins to deviate from its equilibrium value. At that time, $n_3$ is so
small that $K_3$ production in residual DM annihilation comes into equilibrium
with $K_3$ decay.

\begin{figure*}
  \begin{tabular}{c@{\quad}c}
    \includegraphics[width=0.48\textwidth]{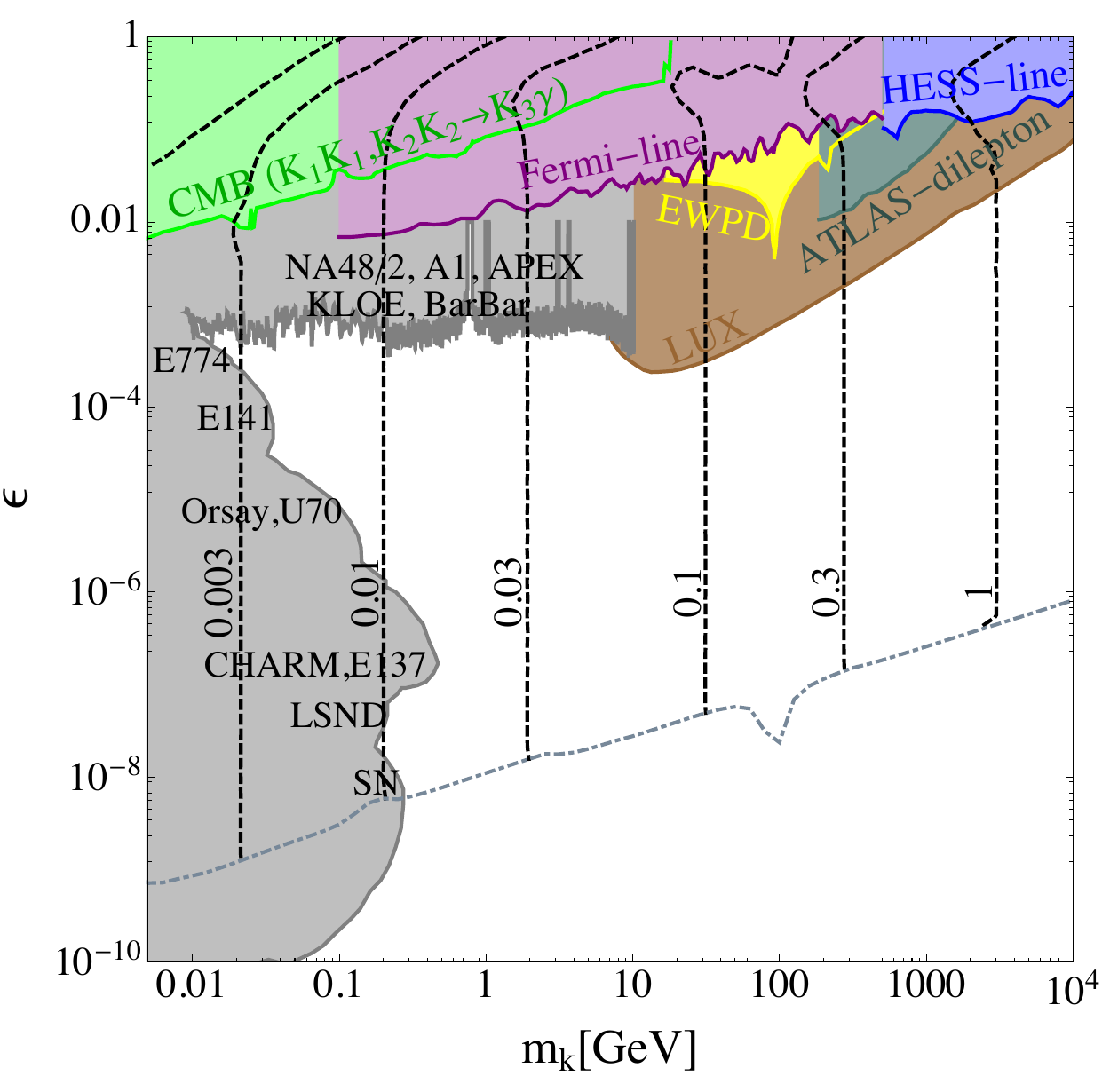} &
    \includegraphics[width=0.48\textwidth]{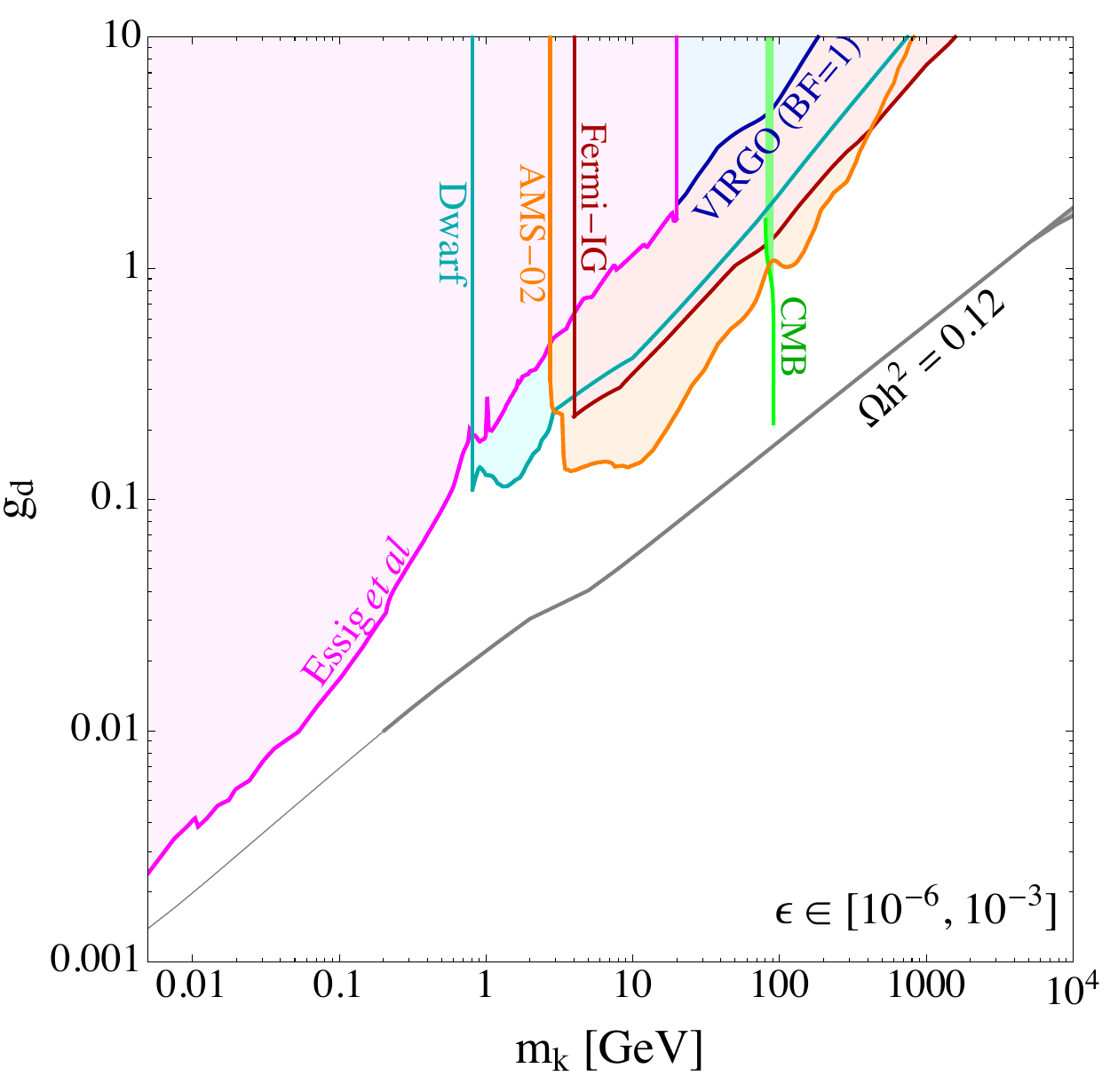} \\
    (a) & (b)
  \end{tabular}
  \caption{(a): Parameter space for the dark $SU(2)_d$ model of Impeded
    DM as a function of DM mass $m_k$ and kinetic mixing parameter $\eps$.
    Dashed black lines indicate, for each combination of $m_k$ and $\eps$,
    the value of the dark sector gauge coupling $g_d$ required to obtain the
    correct DM relic density. Shaded regions show constraints from direct
    detection (LUX, brown)~\cite{LUX2016, Akerib:2015rjg} (see also PandaX-II~\cite{Tan:2016zwf}), the CMB
    (light green)~\cite{Ade:2015xua}, gamma ray line searches in Fermi-LAT
    (purple)~\cite{Ackermann:2013uma} and H.E.S.S. (blue)~\cite{Ovanesyan:2014fwa},
    collider searches (ATLAS dilepton, dark green)~\cite{Cline:2014dwa},
    electroweak precision data (EWPD, yellow)~\cite{Hook:2010tw}, and dark
    photon searches
    (gray)~\cite{Bergsma:1985is,Konaka:1986cb,Riordan:1987aw,Bjorken:1988as,Bross:1989mp,
      Davier:1989wz,Athanassopoulos:1997er,Astier:2001ck,Adler:2004hp,Bjorken:2009mm,
      Artamonov:2009sz,
      Essig:2010gu,Blumlein:2011mv,Gninenko:2012eq,Blumlein:2013cua,Abrahamyan:2011gv,
      Merkel:2014avp,Merkel:2011ze,Aubert:2009cp,Curtin:2013fra,Lees:2014xha,Bernardi:1985ny,
      MeijerDrees:1992kd,Archilli:2011zc,Gninenko:2011uv,Babusci:2012cr,Adlarson:2013eza,
    Agakishiev:2013fwl,Adare:2014mgk,Batley:2015lha,KLOE:2016lwm}.
    In the region below the dot-dashed gray line, $\Gamma_{K_3} < \text{H}(T=m_k)$
    and the model is in the ``co-decaying'' regime~\cite{Dror:2016rxc}.
    Constraints labeled with ``-line" correspond to bounds on a monoenergetic
    gamma ray flux from
    $K_{1} K_{1}, K_{2} K_{2} \to K_3 \gamma$.
    (b): Constraints in the $m_k$ vs.\ $g_d$ plane in
    the region $10^{-6} \lesssim \eps \lesssim 10^{-3}$, in which the DM
    relic density is independent of $\eps$.  In the gray band, the correct
    density is obtained.
    We compare to constraints from Fermi-LAT gamma ray searches in dwarf
    galaxies (cyan) \cite{Ackermann:2015zua, Elor:2015bho},
    in the Virgo cluster (dark blue)~\cite{Ackermann:2015fdi}, and
    in the inner Milky Way (dark red) \cite{Massari:2015xea}, to
    exclusion limits from AMS-02 positron data (orange),
    and to a combination of x-ray and gamma-ray bounds from a compilation
    by Essig et al. (magenta)~\cite{Essig:2013goa}.
    We plot the CMB constraint for a narrow window $m_k \in \left[
    m_{W,\text{SM}}, m_{Z,\text{SM}} \right]$
    where $\Delta > 0$ and use $\eps = 10^{-3}$ for this constraint.}
  \label{fig:constraints}
\end{figure*}

We study the parameter dependence of the DM relic density in
\cref{fig:constraints}.  The left panel in this figure shows the value of the
$SU(2)_d$ gauge coupling $g_d$ required to obtain the correct DM relic density
as a function of the DM mass $m_k$ and the kinetic mixing parameter $\eps$
(dashed black contours).  We see that for large DM mass larger $g_d$ is
required to compensate for smaller annihilation cross-sections.  At $10^{-6}
\lesssim \eps \lesssim 10^{-2}$, the relic density is independent of $\eps$
because the dominant annihilation process in this regime, $K_{1} K_{1}, K_{2} K_{2} \to
K_3 K_3$ happens entirely in the dark sector. At $\eps \gtrsim 10^{-2}$,
the $\eps^2$-suppressed annihilation channels to $K_3 \gamma$, $f \bar{f}$, and
$W^+ W^-$ also become significant, leading to distortion of the contours in
\cref{fig:constraints} (a).  At very small $\eps$, on the other hand, the
$K_3$ decay rate drops below the Hubble rate. The lingering $K_3$ can
annihilate back to $K_{1,2} K_{1,2}$, thus reducing the net DM annihilation
rate and delaying freeze-out unless $g_d$ is increased or $m_k$ is lowered.  In
this regime, ``cannibal processes'' such as $K_1 K_1 K_2 \to K_1 K_3$
need to be taken into account \cite{Dror:2016rxc}.
We also note the small dip in the
$g_d = \text{const}$ contours in \cref{fig:constraints} (a) around $m_k =
m_Z$.  In this region, the mixing between $K_3$ and the $Z$ is large even for
very small $\eps$.

\subsection{Direct Detection}
\label{sec:dd}

Direct detection of Impeded DM in the
$SU(2)_d$ model is complementary to indirect searches as it is sensitive to the
$K_1 K_2 f \bar{f}$ coupling, which does not contribute significantly to DM
annihilation, being velocity and $\eps^2$ suppressed (see \cref{tab:classification}).
The relevant processes
for direct detection are the spin-independent $t$-channel reactions $K_{1,2} q
\to K_{2,1} q$, mediated by $\gamma$, $Z$, and $K^3$. Since the typical
momentum transferred in DM--nucleus scattering is $\ll m_k,\,m_Z$, the photon
mediated diagram dominates over the $Z$ and $K_3$ mediated diagrams, so we
neglect the latter.  In principle, the dark Higgs boson $\phi$ can also mediate
DM--nucleus scattering via mixing with the SM Higgs through the Higgs portal,
\cref{eq:L-H-portal}.  However, the corresponding amplitudes are suppressed by
the mass of $\phi$, the small Yukawa couplings of the Higgs, and by our
assumption that Higgs mixing is tiny. Therefore, we neglect $\phi$-mediated
scattering here.

The spin-independent (SI) DM--nucleus scattering cross-section in the $SU(2)_d$
model is
\begin{align}
  \frac{d\sigma_\text{SI}}{dE_r}
    &= \frac{2\pi \alpha_\text{em} \alpha_d (Z \varepsilon )^2}
            {3 m_k^2 E_r}
            \biggl( 1 + \frac{E_r}{4 E_\text{in}}
                        \frac{m_N^2 - 2 m_k m_N - m_k^2}{m_k m_N} \biggr)
            F_\text{SI}^2(E_r) \,,
  \label{eq:su2dd}
\end{align}
where $E_r$ is the nuclear recoil energy, $Z$ is the nuclear charge,
$\alpha_\text{em} \equiv e^2/(4\pi)$ and $\alpha_d \equiv g_d^2/(4\pi)$ are the
electromagnetic and $SU(2)_d$ fine structure constants, respectively,
$E_\text{in} = m_k v^2_\text{in} / 2$ is the kinetic energy of the incoming DM
particle, and $F_\text{SI}(E_r)$ is the nuclear form
factor~\cite{Jungman:1995df}.  To obtain the spin-independent result
in \cref{eq:su2dd}, we have computed
the scattering of $K_1$, $K_2$ on a scalar particle of charge $Z$, thus
neglecting the nuclear spin. Spin-dependent scattering exists as well,
but as usual constraints are much weaker as the $Z^2$ enhancement is absent.
Note that \cref{eq:su2dd} has some similarity
with the scattering cross section for dipolar dark matter~\cite{Masso:2009mu,
DelNobile:2012tx}. This is not surprising as we argued in \cref{sec:SU2model}
that the $K_1 K_2 \gamma$ coupling is in fact a magnetic dipole coupling.

To calculate direct detection constraints, we use data from the LUX
experiment corresponding to 332 live days~\cite{Akerib:2016vxi}, (see also
PandaX-II results~\cite{Tan:2016zwf} from first 98.7-day data, which has
comparable limit to LUX).
The LUX constraint is presented in ref.~\cite{Akerib:2016vxi}
as a
mass-dependent limit on the total DM--nucleon scattering cross-section
$\sigma_n$, assuming the latter to be independent of the DM velocity.
This
assumption is violated for the photon-mediated scattering processes relevant in our
$SU(2)_d$ model. Therefore, we first compute $\sigma_n$ in a contact operator
model with a fermionic DM candidate $\chi$, for instance $\mathcal{L} \supset
\bar\chi \gamma_\mu \chi \bar{q} \gamma^\mu q$ and choose the coupling such
that the LUX limit is saturated.  We then compute the differential event rate
$dR/dE_r$ for this operator, taking into account the Maxwell--Boltzmann-like DM
velocity distribution, and multiply by the efficiency for nuclear recoil events
in LUX~\cite{Akerib:2015rjg}.
We integrate $dR/dE_r$ over the energy
range $\text{1.1\,keV} < E_r < \text{100\,keV}$ to obtain the maximum total
number of events $N^\text{max}$ consistent with LUX data.  We then compute
$\int dE_r \, dR/dE_r$ also in our model. By requiring the result to match
$N^\text{max}$ determined for the contact operator, we obtain a constraint on
the coupling $g_d$.  This constraint is shown in \cref{fig:constraints} (a)
in brown.  We see that it is stronger than indirect bounds and collider bounds
for DM masses between 10\,GeV and 10\,TeV.

\subsection{Constraints from the Cosmic Microwave Background}
\label{sec:cmb}

Another important constraint on any model in which DM can annihilate arises
from observations of the cosmic microwave background (CMB). In particular,
the extra energy injected into the primordial plasma due to DM annihilation
would delay recombination and thus leave observable imprints in the CMB
\cite{Adams:1998nr, Padmanabhan:2005es, Galli:2009zc, Slatyer:2009yq}.
The impact of DM on the CMB
is characterized by the ``energy deposition yield''~\cite{Finkbeiner:2011dx, Madhavacheril:2013cna}
\begin{align}
  p_\text{ann} = f_\text{eff} \frac{\ev{\sigma v}}{m_\text{DM}} \,.
\end{align}
Here, $f_\text{eff}$ gives the efficiency with which the energy released in DM
annihilation is absorbed by the primordial plasma.

For the specific case of the $SU(2)_d$ model, we need to consider the
annihilation processes shown in \cref{tab:classification}. As in the previous
sections, we neglect DM annihilation to $\phi\gamma$, assuming that $\phi$ is
sufficiently heavy for this channel to be closed.  We also note that
annihilation via $K_{1} K_{2} \to f\bar{f}, W^+ W^-$ is subdominant at the CMB
epoch because of $v_\text{rel}^2$ and $\eps^2$ suppression, as is
$K_{1} K_{1}, K_{2} K_{2} \to \gamma\gamma$ because of $\eps^4$ suppression.

The annihilation cross-section
for $K_{1} K_{1}, K_{2} K_{2} \to K_3 K_3$ is phase space suppressed by the
factor $\sqrt{v_\text{rel}^2/4 - 2\Delta/m_k}$, therefore we need to estimate
the DM velocity at the time of CMB decoupling.  To do so, we need to determine
the temperature at which DM kinetically decouples from the SM, i.e.\ the
temperature at which $K_{1,2} f \to K_{2,1} f$ scattering freezes out.
(Scattering of $K_{1,2}$ on photons via $t$-channel $K_{1,2}$ exchange is
negligible as the cross-section is proportional to $\eps^4$.) It turns out
that, in most of the parameter space considered here, this happens no later
than at $T \gtrsim 1\,\text{MeV}$, when $e^+ e^-$ annihilation reduces the
density of SM fermions by $\sim 10$ orders of magnitude.  Afterwards, the
kinetic energy of DM drops quickly as $a^{-2}$, where $a$ is the scale factor
of the Universe. Therefore, by the time of recombination, the dark sector
temperature has dropped to $\lesssim 10^{-6}\,\text{eV}$.
We conclude that, at the CMB epoch the DM temperature is typically too
low to overcome the mass splitting
$|\Delta| \sim m_k \eps^2$ in the process $K_{1} K_{1}, K_{2} K_{2} \to K_3 K_3$,
except at very small $\eps$ and in a small
mass window with $m_{W,\text{SM}} < m_k < m_{Z,\text{SM}}$ where $\Delta > 0$
(see \cref{eq:su2-delta}). We plot the CMB constraint from
$K_{1} K_{1}, K_{2} K_{2} \to K_3 K_3$ in this narrow window,
for $\eps = 10^{-3}$ in \cref{fig:constraints} (b). We see that
the resulting limit is $g_d \lesssim 0.2 (10^{-3}/\eps)^{1/4}$.

Finally, we need to consider the annihilation process $K_{1} K_{1}, K_{2} K_{2}
\to K_3 \gamma$. For this annihilation channel, $f_\text{eff}$ can be written as
\begin{align}
  f_\text{eff} = \frac{E_{K_3}}{E_{K_3}+E_\gamma} f_\text{eff}^{K_3}(E_{K_3})
    + \frac{E_{\gamma}}{E_{K_3}+E_\gamma} f_\text{eff}^\gamma(E_\gamma) \,,
  \label{eq:feff}
\end{align}
where $E_{K_3} \thickapprox \frac{5}{4}m_k$ and $E_{\gamma} \thickapprox \frac{3}{4}m_k$.
In \cref{eq:feff}, the contributions to $f_\text{eff}$ from $K_3$ ($f_\text{eff}^{K_3}$)
and from photons ($f_\text{eff}^\gamma$) is weighted by respective energy fraction
because the CMB is sensitive to energy injection into the primordial plasma. $f_\text{eff}^{K_3}$ is given by
\begin{align}
  f_\text{eff}^{K_3}(E_{K_3}, m_k, \eps)
    \thickapprox \sum_i \BR_{K_3 \to SM_i SM_i}(m_k,\eps) f_\text{eff}^{SM_i SM_i}(E_{K_3}/2) \,.
  \label{eq:K3eff}
\end{align}
Here, the sum runs over all SM final states into which $K_3$ can decay, and
$f_\text{eff}^{SM_i SM_i}$ are the corresponding efficiency factors for each
final state. We take these, as well as $f_\text{eff}^\gamma$ from
ref.~\cite{Slatyer:2015jla}.  We make the approximation here that the energy of
each $SM_i$ particle is $E_{K_3}/2$ in the laboratory frame. Their actual
energy is distributed around $E_{K_3}/2$, but since the energy of $K_3$ is very
close to its mass $m_k$, the distribution is very close to a delta function.
Moreover, $f_\text{eff}^{SM_i SM_i}$ changes only mildly with $E_{SM_i}$,
therefore our assumption is reasonable.  For $m_k$ smaller than the QCD scale,
the calculation of $f_\text{eff}^{K_3}$ follows the procedure from
ref.~\cite{Liu:2014cma}.  Demanding $p_\text{ann} < 4.1 \times
10^{-28}\,\text{cm}^3\;\text{s}^{-1}\;\text{GeV}^{-1}$~\cite{Ade:2015xua}, we
obtain the constraints shown in green in \cref{fig:constraints} (a).  We see
that CMB constraints from $K_{1} K_{1}, K_{2} K_{2} \to K_3 \gamma$ are
particularly strong at low DM mass, where the annihilation cross-section is
large.  In conventional WIMP models, they exclude thermal relic DM lighter than
$\sim 10$~GeV, while in our $SU(2)_d$ model, they can always be avoided by
choosing $\eps \lesssim 10^{-2}$, a condition that is imposed anyway by dark
photon searches (gray region in \cref{fig:constraints} (a)).

\subsection{Indirect Detection}
\label{sec:indirectD}

In this section, we will investigate indirect astrophysical constraints
on Impeded DM in the $SU(2)_d$ model, in particular from searches for anomalous
signals in continuum gamma rays, charged cosmic rays, and gamma ray lines.

The differential flux of continuum photons from a solid angle interval
$d\Omega$ is
\begin{align}
  \frac{d\Phi}{dE_\gamma d\Omega}
  &= \frac{1}{8 \pi \, c \, m_\text{DM}^2} J(\theta,\phi) \,
     \sum_X \ev{\sigma v}_X  \frac{dN_\gamma^X}{dE_\gamma} \,,
  \label{eq:phi-id}
\end{align}
where $\ev{\sigma v}_X$ is the thermally averaged annihilation cross-section
for a process $X$, $dN_\gamma^X / dE_\gamma$ is the differential
photon spectrum for a single annihilation reaction, and the sum runs over
all accessible final states. The factor $c$ is a symmetry factor, which
is $c=4$ for vector DM. It would be $c=1$ ($c=2$) if DM was a Majorana
(Dirac) fermion.\footnote{While the symmetry factor is different for
different types of DM, the thermal relic cross-sections for different
candidates are modified by an identical factor, so that the expected
gamma ray flux is independent of $c$.}  The factor $J(\theta,\phi)$ in \cref{eq:phi-id}
is the integral over the squared DM density along the line of sight (l.o.s.)
oriented in direction $(\theta,\phi)$. It is given by
\begin{align}
  J(\theta,\phi) = \int_\text{l.o.s.} \! ds \, \rho_\text{DM}^2(s,\theta,\phi) \,.
\end{align}
We describe $\rho_\text{DM}$ as an NFW profile with a local DM density
$\sim 0.3$\,GeV/cm$^3$~\cite{Iocco:2011jz,Catena:2009mf,Salucci:2010qr}, and a scale radius of 20 kpc.
The cosmic ray $e^+$ and $e^-$ spectra are
obtained from an expression analogous to \cref{eq:phi-id}, replacing
$dN_\gamma^X / dE_\gamma$ by the corresponding spectra $dN_{e^\pm}^X /
dE_{e^\pm}$.

The dominant contribution to continuum gamma ray and charged cosmic ray signals
in the $SU(2)_d$ model comes from the annihilation channel $K_{1} K_{1}, K_{2} K_{2}\to
K_3 K_3$. Even though we have seen above that this process is kinematically
forbidden at the CMB epoch, it opens up again later, when DM particles are
reaccelerated as they fall into the gravitational potential wells of newly
forming galaxies and clusters. Observable signals arise from $K_{1} K_{1}, K_{2} K_{2} \to
K_3 K_3$ when $K_3$ decays to SM particles through its kinetic mixing with the
photon and the $Z$.  These decays contribute to cosmic $e^+$ and $e^-$ fluxes
through $K_3 \to e^+ e^-$, and to $e^+$, $e^-$, and to gamma ray fluxes through
final state radiation and $K_3 \to \text{mesons}$, followed by meson decays.
For $m_k \lesssim 3$\,GeV, we compute the spectra
$dN_{e^\pm,\gamma}/dE_{e^\pm,\gamma}$ from $e^+ e^- \to \text{hadrons}$ data
following ref.~\cite{Liu:2014cma}. At larger $m_k$, we compute the $K_3$ decay
rates to quark and lepton pairs and then use ref.~\cite{Cirelli:2010xx} to
obtain the resulting cosmic ray spectra.

The high-energy $e^+$ and $e^-$ can also upscatter ambient photons to gamma-ray
energies via inverse Compton scattering (ICS), providing an additional secondary
contribution to the gamma-ray flux. This contribution depends on the propagation
of the charged particles, and so has additional uncertainties relative to the prompt
photon emission from annihilation. For the constraints we discuss below, only
those from Fermi observations of the Virgo cluster include the ICS component.

We also consider DM annihilation to final states containing mono-energetic photons,
where the dominant signal channel is $K_{1} K_{1}, K_{2} K_{2} \to K_3 \gamma$. In this
case $dN^X_\gamma / dE_\gamma$ is just a $\delta$ function. As in the previous
sections, we do not consider $K_{1} K_{2} \to \phi \gamma$, assuming
this channel to be kinematically forbidden.

We compare the predicted cosmic ray spectra to the following data sets
\begin{itemize}
  \item {\bf Fermi-LAT observations of dwarf galaxies.}
    We use the bin-by-bin likelihood provided by the Fermi-LAT
    collaboration~\cite{Ackermann:2015zua}, based on observations of 15
    non-overlapping dwarf galaxies.  Using \cref{eq:1133}, we can translate
    this likelihood into limits on the annihilation cross-section $\ev{\sigma
    v_\text{rel}}_{11 \to 33}$ and hence $g_d$.  In computing $\ev{\sigma
    v_\text{rel}}_{11 \to 33}$, we account for the different root mean
    square (rms) velocity $v_0$ in each dwarf galaxy, and we use
    $\ev{v_\text{rel}} = (2/\sqrt{\pi}) v_0$.  This approach is valid in
    the regime where the cross-section is linearly dependent on velocity;
    in the forbidden regime where $\varepsilon > v_0$, it may mis-estimate
    $\langle \sigma v_\text{rel} \rangle$ (since in this case the
    cross-section will be sensitive to the high-velocity tail of the
    velocity distribution), but in this regime the cross-section will in
    any case be very small.

  \item {\bf Fermi-LAT observations of the Virgo cluster.} This constraint is
    based on three years of Fermi-LAT data, presented in
    ref.~\cite{Ackermann:2015fdi} as upper limits on $\ev{\sigma
    v_\text{rel}}_{\text{SM}_i \text{SM}_i}(m_\text{DM})$, the thermally
    averaged DM annihilation cross-section into different final states
    consisting of pairs of SM particles $\text{SM}_i$.  We impose that
    \begin{align}
      \frac{2}{c} \Big[\ev{\sigma v_\text{rel}}_{11 \to 33}
                     + \ev{\sigma v_\text{rel}}_{22 \to 33} \Big]
      \Big[
        (\BR_{K_3 \to \text{SM}_i \text{SM}_i})^2
      + \BR_{K_3 \to \text{SM}_i \text{SM}_i}
          \big( 1 - \BR_{K_3 \to \text{SM}_i \text{SM}_i} \big)
      \Big]
    \end{align}
    should be below the limiting value of
    $\ev{\sigma v_\text{rel}}_{\text{SM}_i \text{SM}_i}(m_k/2)$. Here, $c$ is the same
    symmetry factor as in \cref{eq:phi-id}, and the last term describes the average
    number of $K_3$ decays to $\text{SM}_i \text{SM}_i$.
    In computing
    $\ev{\sigma v_\text{rel}}_{11 \to 33}$ and $\ev{\sigma v_\text{rel}}_{22 \to 33}$,
    we use the rms velocity of the Virgo
    cluster, $v_0=525\,\text{km}/\text{sec}$, and set again $\ev{v_\text{rel}} =
    (2/\sqrt{\pi}) v_0$. We find that the most constraining $K_3$
    decay modes are $\tau^+ \tau^-$ at $m_k \lesssim 40 \text{GeV}$, $b\bar{b}$
    at intermediate $m_k \in \left[40, 200 \right] \text{GeV}$, and $e^+ e^-$ at
    $m_k \gtrsim 200 \text{GeV}$.
    The strong constraint on annihilation to $e^+ e^-$ at high masses arises from
    inverse Compton scattering of the electrons on the CMB, which produces photons
    in the Fermi-LAT energy range.
    Note that the authors of \cite{Ackermann:2015fdi} multiply the DM annihilation
    cross-section by a boost factor to account for enhanced annihilation
    in overdense DM subhalos. We do not include boost
    factors here because (a) the size of this boost factor is highly uncertain,
    so constraints assuming a large boost factor are difficult to make robust,
    and (b) since the rms velocity in DM subhalos is much lower than in
    the host halo, $\ev{\sigma v_\text{rel}}_{11 \to 33}$ and
    $\ev{\sigma v_\text{rel}}_{22 \to 33}$ will be lower
    for DM particles bound in subhalos, especially for the very small subhalos
    that typically contribute much of the boost.

  \item {\bf Gamma ray constraints from the inner Milky Way.}
    These limits are derived in analogy to the Virgo limits,
    but based on the results of ref.~\cite{Massari:2015xea}, assuming an NFW
    profile for the Milky Way.
    We assume an rms velocity $v_0 = 220\,\text{km}/\text{sec}$ for the Milky Way,
    but we remind the reader that the velocity dispersion in the Galactic Center
    region is highly uncertain, see for instance \cite{Iocco:2015xga}.

  \item {\bf Combined x-ray, gamma ray, and $e^+e^-$ limits for light DM.}
    For low mass DM ($1\,\text{MeV} \lesssim m_k \lesssim 10\,\text{GeV}$),
    Essig \textit{et al.} \cite{Essig:2013goa} have compiled x-ray and gamma ray
    constraints for the annihilation channel $\text{DM}\,\text{DM} \to e^+ e^-$.
    They use data from the HEAO-1~\cite{Gruber:1999yr}, INTEGRAL~\cite{Bouchet:2008rp},
    COMPTEL~\cite{COMPTEL}, EGRET~\cite{Strong:2004de}, and Fermi~\cite{Ackermann:2012pya}
    satellites. We translate these limits into bounds on $g_d$ in the same way
    as for Fermi-LAT limits from the Virgo cluster and the Milky Way.

  \item {\bf AMS-02 data on $e^+$ and $e^-$ fluxes.}
    Monoenergetic $e^+ e^-$ pairs produced in $K_3$ decays can generate
    bump-like features in the cosmic electron and positron fluxes observed by
    AMS-02.  We use in particular the AMS-02 measurement of the positron flux
    \cite{Aguilar:2014mma} and follow the approach of ref.~\cite{Elor:2015bho}
    to derive a bound on
    $\ev{\sigma v_\text{rel}}_{33}$ from it. In computing $\ev{\sigma
    v_\text{rel}}_{33}$, we assume $v_0 = 220\,\text{km}/\text{sec}$.  Note
    that our bound is more conservative than the one from
    ref.~\cite{Bergstrom:2013jra} since we assume larger magnetic fields in
    simulating $e^+ e^-$ propagation~\cite{Evoli:2008dv,2011ascl.soft06011M}.

  \item {\bf Gamma ray line searches in Fermi-LAT and H.E.S.S.}
    Even though the cross-section for $K_{1} K_{1}, K_{2} K_{2} \to K_3 \gamma$
    is suppressed by $\eps^2$, we expect competitive limits from these
    channels thanks to the cleanliness of gamma ray line signatures.
    We derive these limits using the data from ref.~\cite{Ackermann:2013uma,
    Ovanesyan:2014fwa} {}\footnote{Ref.~\cite{Ovanesyan:2014fwa} has translated
    the H.E.S.S. limits~\cite{Abramowski:2013ax} from Einasto profile
    to NFW profile. }, (see also \cite{Profumo:2016idl}).
\end{itemize}
The above results are summarized in \cref{fig:constraints} (b). We see that
AMS-02 provides the most stringent constraints for $3\,\text{GeV} \lesssim m_k
\lesssim 400\,\text{GeV}$, followed by gamma ray constraints from the inner
galaxy.
The constraints from Fermi-LAT gamma ray searches
in dwarf galaxies provide the strongest bound for DM masses around a
few GeV, where the annihilation products mostly lie below the energy
threshold of AMS-02. The dwarf bounds are only a factor of
few weaker than those from AMS-02 and Fermi observations of the inner
Galaxy over the remainder of the mass range, and have smaller
systematic uncertainties. At DM masses below 1\,GeV, the dominant
decay channel of $K_3$ is $e^+e^-$.
In this case, the strongest constraint arises from limits on x-ray and gamma ray photons produced as final state radiation from DM annihilation in the Milky Way.
Galactic observations offer on the one hand
large statistics, and on the other hand large DM velocities,
which is important
for the $v_\text{rel}$-suppressed annihilation channel
$K_1 K_1, K_2 K_2 \rightarrow K_3 K_3$.
Note that when $\eps \gtrsim 10^{-3}$ (above the range assumed in
\cref{fig:constraints} (b), and in tension with other limits according to
\cref{fig:constraints} (a)), $|\Delta|$ is large and shuts off
the $K_{1} K_{1}, K_{2} K_{2}  \to K_3 K_3$ channel, see \cref{eq:1133}.
This happens first in dwarf galaxies, where $v_0$ is lowest.

\section{Dark Pions as Impeded DM with $\Delta > 0$ }
\label{sec:dark-pion}

\subsection{Model}
\label{sec:compositemodel}

We now switch gears and discuss a second realization of Impeded DM in a
concrete model. In particular, we introduce a composite hidden sector based on
an $SU(N) \times U(1)'$ gauge symmetry, analogous to the strong and
electromagnetic interactions of the SM (see refs.~\cite{Ryttov:2008xe,
  Bai:2010mn, Bai:2010qg, Hur:2011sv, Fan:2012gr, Frigerio:2012uc,
  Buckley:2012ky, Bhattacharya:2013kma, Holthausen:2013ota, Cline:2013zca, Hochberg:2014dra,
  Hochberg:2014kqa,Yamanaka:2014pva, Carmona:2015haa, Ametani:2015jla, Hochberg:2015vrg,
Hatanaka:2016rek} for similar models). We assume the existence of two species
of light ``dark quarks'' $u_d$ and $d_d$ with the charge assignments listed in
\cref{tab:fieldcontent}. We also introduce a dark scalar field $\phi$ that
breaks $U(1)'$ by two units, giving mass to the dark photon.
In analogy to QCD, the global chiral symmetry of the
dark quarks is broken at energies below the strong coupling scale $\Lambda_N$.
The associated Nambu--Goldstone bosons (dark pions), $\pi^+_d$, $\pi^-_d$
constitute excellent DM candidates, stabilized by a $Z_2$ symmetry,
a residual of the broken dark $U(1)'$ symmetry.
(Note that the superscripts here
refer to the $U(1)'$ charge of the dark pions, not an electromagnetic charge.)
Their neutral partner, $\pi^0_d$, is unstable
and can decay through the chiral anomaly to dark photons.

In the broken phase of chiral symmetry, the effective Lagrangian of the
model is~\cite{Fan:2012gr}
\begin{align}
  \mathcal{L}
    = \frac{1}{4} f_\pi^2 \Tr \big[ \partial_\mu U^\dagger \partial^\mu U \big]
    +  \mu \frac{f_\pi^2}{2} \Tr \big[  U^\dagger M + M^\dagger U \big] \,,
\end{align}
where we will refer to $f_\pi$ as the dark pion decay constant (even though the
dark $\pi_d^\pm$ are stable).
The matrix $U$ is defined as $U \equiv \exp(i\pi^a_d \sigma^a / f_\pi)$ with
the Pauli matrices $\sigma^a$,
$M$ is the $2 \times 2$ mass matrix for $u_d$ and $d_d$, and
\begin{equation}
  \pi^a_d \sigma^a =
  \begin{pmatrix}
    \pi^0_d          & \sqrt{2} \pi^+_d \\
    \sqrt{2} \pi^-_d & -\pi^0_d
  \end{pmatrix} \,.
\end{equation}
Note that the mass matrix $M$ is diagonal; since $\phi$ carries two units
of $U(1)'$ charge, it cannot induce mixing between $u_d$ and $d_d$ even after
breaking $U(1)'$.

\begin{table}
  \begin{center}
    \begin{minipage}{5cm}
      \begin{ruledtabular}
      \begin{tabular}{ccc}
        & SU(N) &  $U(1)^\prime$    \\
        \hline
        $u_d$  & $\square$ & $2/3$  \\
        $d_d$  & $\square$ & $-1/3$ \\
        $\phi$ &      1    & 2
      \end{tabular}
      \end{ruledtabular}
    \end{minipage}
  \end{center}
  \caption{Field content and quantum numbers of the dark pion model,
    where $\square$ stands for the fundamental representation of the dark $SU(N)$.
    We show here only the field content necessary for the Impeded DM
    phenomenology, but it is important to keep in mind that additional particles
    like heavy dark leptons are necessary for anomaly cancellation.}
  \label{tab:fieldcontent}
\end{table}

Dark pion DM can behave as Impeded DM if there is a mass splitting between
$\pi^\pm_d$ and $\pi^0_d$.  Such a mass splitting could have two different
origins: different $u_d$ and $d_d$ masses, $m_{u_d} \neq m_{d_d}$, and $U(1)'$
radiative corrections. We assume for simplicity that $u_d$ and $d_d$ are
degenerate in mass, i.e.\ that dark isospin is unbroken.
This means in particular that we neglect the $\eta^0_d$ meson, which could mix
with $\pi^0_d$ if $m_{u_d} \neq m_{d_d}$. Such mixing would be proportional
to $m_{u_d} - m_{d_d}$ and would give a mass splitting between $\pi^\pm_d$
and $\pi^0_d$ of order
$(m_{u_d} - m_{d_d})^2 m_{\pi^\pm_d}^3 / [2 (m_{u_d} + m_{d_d})^2
m^2_{\eta_d^0} ]$~\cite{Fan:2012gr}.

The mass splitting between $\pi^\pm_d$ and $\pi^0_d$ is then obtained
from the self-energy diagrams of $\pi^\pm_d$ through $A'$. For light $A'$,
the mass splitting is estimated to be~\cite{Harigaya:2015ezk, Bai:2015nbs,
Das:1967it, Donoghue:1996zn},
\begin{align}
  m_{\pi^\pm_d}^2 - m_{\pi^0_d}^2 &\approx  \frac{g'^2}{16 \pi^2} \Lambda^2_N\\
  \Delta \equiv   m_{\pi^\pm_d} - m_{\pi^0_d}
         &\approx \frac{g'^2}{16 \pi^2} \frac{\Lambda^2_N}{2 m_\pi} \,.
   \label{eq:pisplitting}
\end{align}
In the following, we use the value $\Lambda_{N} = 4 \pi f_\pi$
for the dark sector confinement scale.  Note that $\Delta$ in the
dark pion model is always positive, i.e.\ DM is always
heavier than its annihilation product $\pi^0$. Thus, annihilation is
never kinematically forbidden.

In the following, we will also need the rate of the anomaly-mediated decay
$\pi^0_d \to A' A'$, which we calculate to be
\begin{equation}
  \Gamma(\pi^0_d \to A' A')
    = \frac{g'^4 m_\pi^3}{1024 \pi^5 f_\pi^2}
      \bigg(1 - \frac{4 m_A'^2}{m_\pi^2} \bigg)^{3/2} \,.
\end{equation}

\subsection{Constraints from relic abundance, direct and indirect detection}
\label{sec:darkpion-constraints}

Annihilation of the DM particles $\pi^\pm_d$ in the dark pion scenario is dominated
by the process $\pi^+_d \pi^-_d \to \pi^0_d \pi^0_d$. The amplitude for this
reaction is $\mathcal{M}(\pi^+_d \pi^-_d \to \pi^0_d \pi^0_d) = (s - m_\pi^2) /
f_{\pi}^2$ \cite{Weinberg:1966kf}, where $m_\pi \equiv m_{\pi^\pm_d}$ is the dark pion mass.
The cross-section is then given by
\begin{align}
  (\sigma v_\text{rel})_{00}
    &= \sigma_0 \times \sqrt{\frac{v_\text{rel}^2}{4} + \frac{2 \Delta}{m_\pi}}
                                                      \nonumber\\
    &\simeq 6 \times 10^{-26} \text{cm}^3 \, \text{sec}^{-1}
      \times \bigg( \frac{m_\pi/f_\pi^2}{7 \times 10^{-4} \text{GeV}^{-1}} \bigg)^2 \, \quad  \text{at freeze-out,}
  \label{eq:piAnn}
\end{align}
where $\sigma_0 = 9/(64\pi) m_\pi^2 / f_\pi^4$.
The estimate in the second line of \cref{eq:piAnn} is for the time of
DM freeze-out, where $v_\text{rel} \sim 0.47$, and assuming $\Delta / m_\pi \ll 1$.
From the requirement of obtaining the correct thermal relic cross-section,
we then obtain
\begin{align}
  \frac{m_\pi}{f_\pi^2} \sim 7 \times 10^{-4} \, \text{GeV}^{-1} \,.
  \label{eq:mpi-fpi}
\end{align}
In the following, we will use this condition to determine $f_\pi$ as a function
of $m_\pi$.

\begin{figure*}
  \centering
  \begin{tabular}{c}
    \includegraphics[width=0.7\textwidth]{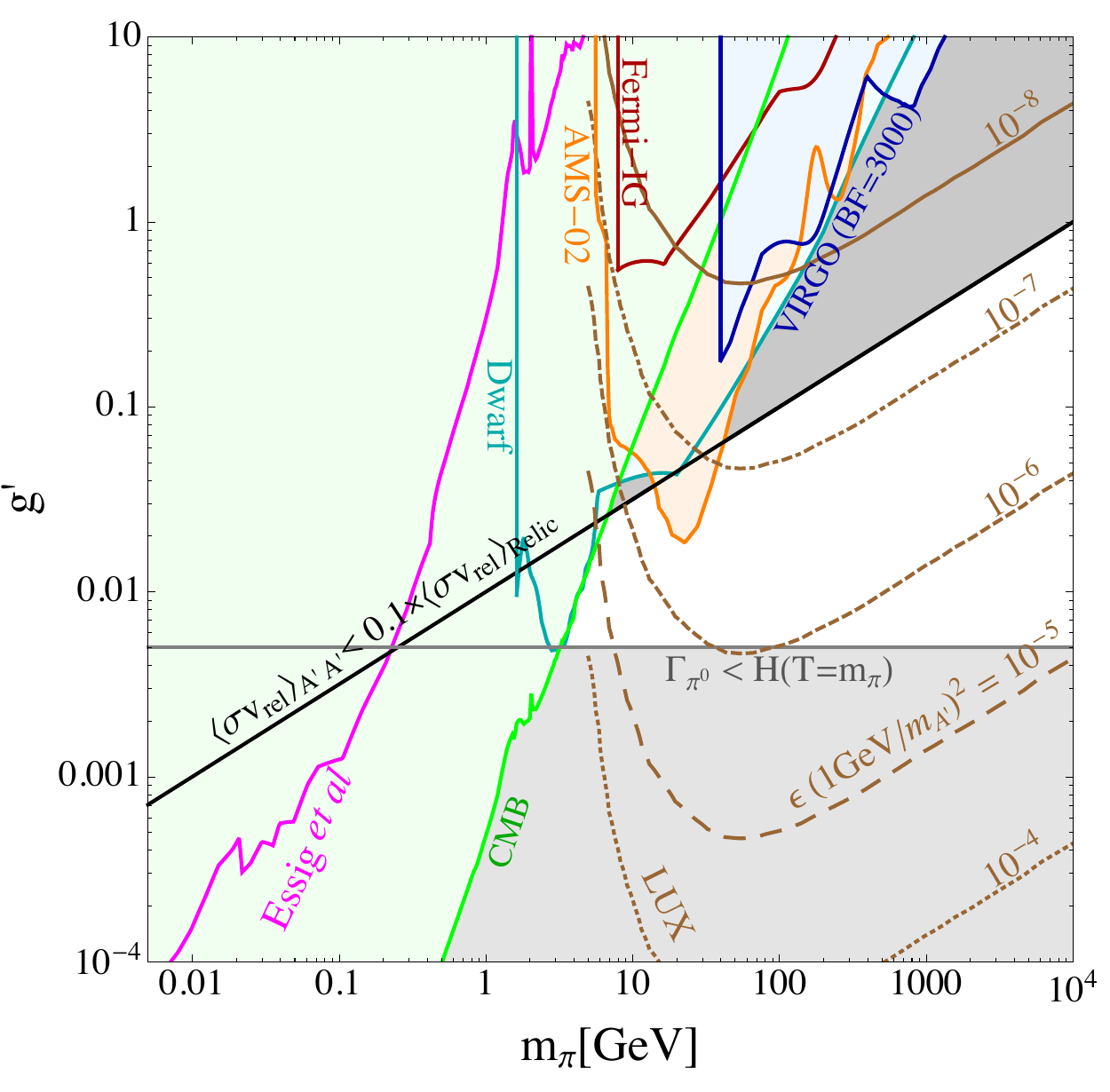}
  \end{tabular}
  \caption{Constraints on the parameter space of the dark pion model from direct
    detection data~\cite{LUX2016, Akerib:2015rjg} and indirect searches.
    The indirect detection constraints are similar to those shown
    in \cref{fig:constraints} (b). We focus on the annihilation process
    $\pi^+_d \pi^-_d \to \pi^0_d \pi^0_d$. For each
    combination of $m_\pi$ and $g'$, the dark pion decay constant $f_\pi$ is
    determined from the relic density requirement \cref{eq:mpi-fpi}.
    In the large-$g'$ region above the diagonal black line, this condition
    is not strictly valid as annihilation via $\pi^+_d \pi^-_d \to A'A'$
    becomes relevant. In the region below the horizontal gray line, the relic
    density is modified by a small $\pi^0_d$ width, preventing $\pi^0_d$ from maintaining
    equilibrium with the SM.}
  \label{fig:constraintpion}
\end{figure*}

DM can annihilate also via $\pi^+_d \pi^-_d \to A' A'$, with cross-section
\begin{align}
  (\sigma v_\text{rel})_{A' A'}
    \simeq \frac{g'^4}{8 \pi m_\pi^2}
           \bigg(1 - \frac{m_{A'}^2}{m_\pi^2} + \frac{3m_{A'}^4}{8m_\pi^4} \bigg)
           \frac{\sqrt{1 - m_{A'}^2/m_\pi^2}}{\big[1 - m_{A'}^2 / (2m_\pi^2)\big]^2} \,.
  \label{eq:pi-ann-A'A'}
\end{align}
In the following, we will neglect this second annihilation channel on the
grounds that the $U(1)'$ gauge coupling $g'$ should be much smaller than the
$m_\pi / f_\pi$ to keep the model QCD-like. Requiring that
$(\sigma v_\text{rel})_{A' A'} < 0.1 (\sigma v_\text{rel})_{00}$ leads to
the requirement $g' \lesssim 0.01 \sqrt{m_\pi / \text{GeV}}$.
In \cref{fig:constraintpion}, where we plot the parameter space of
the dark pion model, this condition is satisfied below the diagonal black line.

To keep $\pi^0_d$ in thermal equilibrium with the SM sector throughout DM
freeze-out, the dark sector should have appreciable interactions with SM
particles. This can be achieved for instance through a kinetic mixing term
of the $A'$,
\begin{align}
  \mathcal{L} \supset \frac{\eps}{2} F_{\mu\nu}' F^{\mu\nu} \,,
  \label{eq:A'-kin-mix}
\end{align}
where $F_{\mu\nu}$ and $F_{\mu\nu}'$ are the field strength tensors of the
photon and the $A'$, respectively.  Requiring the scattering rate for $A' + f
\to \gamma + f$ to be larger than the Hubble rate at freeze-out gives the
constraint
\begin{align}
  \eps \gtrsim \mathcal{O}(10^{-8}) \sqrt{m_\pi/\text{GeV}} \,.
\end{align}
Note that $A'$ decay to $f \bar{f}$ and its inverse are less
efficient than $A' + f \to \gamma + f$
in keeping $A'$ in thermal equilibrium at $\pi^{\pm}$ freeze out
if $m_{A'} \ll m_\pi$. If the $A'$ mass is similar to $m_\pi$,
then $A'$ decay and scattering will have similar efficiency
in keeping $A'$ in equilibrium.
Demanding also that $\pi^0_d$ and $A'$ are in equilibrium through $\pi^0_d
\leftrightarrow A' A'$ leads to the additional requirement
\begin{align}
  g' \gtrsim \mathcal{O}(10^{-3}) (f_\pi^2 m_\pi^{-1} \text{GeV}^{-1})^{1/4}
     \sim 5 \times 10^{-3} \,.
\end{align}
This condition is satisfied above the horizontal gray line in
\cref{fig:constraintpion}.

\vspace{1ex}
In direct detection experiments, dark pion DM can scatter on protons
via $t$-channel $A'$ exchange. The scattering cross-section is
\begin{align}
  \sigma_p
    = \eps^2 e^2 g'^2  \frac{(m_\pi m_p)^2}{\pi m_{A'}^4 (m_\pi + m_p)^2}
    \simeq 10^{-47}\,\text{cm}^2
           \bigg( \frac{g'}{10^{-2}} \bigg)^2
           \bigg( \frac{\eps}{10^{-7}} \bigg)^2
           \bigg( \frac{1\,\text{GeV}}{m_{A'}} \bigg)^4 \,.
  \label{eq:pion-dd}
\end{align}
Based on this expression, we derive constraints on the model parameters from
LUX data~\cite{LUX2016, Akerib:2015rjg}. The result is shown in
\cref{fig:constraintpion} (brown contours) for different values of $\eps
(1\,\text{GeV} / m_{A'})$, as indicated in the plot.

\vspace{1ex}
Dark pion DM is also constrained by indirect astrophysical observations, where
annihilation via $\pi^+_d \pi^-_d \to \pi^0_d \pi^0_d$, followed by $\pi^0_d \to A' A'$
and $A' \to \text{SM} \, \overline{\text{SM}}$ leaves an imprint.
We show the resulting constraints in \cref{fig:constraintpion}.  In this
plot, we have taken $m_{A'} \sim m_{\pi^0_d} / 2$, so that $A'$ particles decay
nearly at rest. Changing the mass of $A'$ will not dramatically change our result.
Constraints are obtained
in the same way as for the $SU(2)_d$ model, see \cref{sec:cmb,sec:indirectD}.

As in the $SU(2)_d$ model, the DM velocity relevant for CMB bounds
is much smaller
than $\sqrt{\Delta/m_\pi}$, so that \cref{eq:piAnn} reduces to
\begin{align}
  (\sigma v_\text{rel})_{00}
    &\simeq 10^{-23} \text{cm}^3 \text{s}^{-1} \frac{g'}{m_\pi/\text{GeV}} \,,
  \label{eq:piannDelta1}
\end{align}
where we have again determined $f_\pi$ from \cref{eq:mpi-fpi}, and $\Delta$ from \cref{eq:pisplitting}.

In fact, for the dark pion model, $v_\text{rel}^2 < \Delta / m_\pi$ holds even
in galaxy clusters as long as $g'$ is not tiny. It holds in particular for $g'$
large enough to keep $\pi^0_d$ in equilibrium in the early Universe, i.e.\
above the horizontal gray line in \cref{fig:constraintpion}. Therefore, we can
always compute the annihilation cross-section using the
$v_\text{rel}$-independent expression in \cref{eq:piannDelta1}.  However, we
now include substructure enhancement in the computation of limits from the
Virgo cluster.  In the plot we have used a substructure boost factor of 3\,000
for the Virgo cluster. Such large boost factors may exist if there is
sufficient small-scale substructure (as discussed in \cite{Ackermann:2015fdi}),
although assuming them could lead to overly stringent constraints. We see,
however, that even for such large boost factors, the limits from Virgo are
superseded by other bounds.

We see from \cref{fig:constraintpion} that constraints from dwarf galaxies
and from AMS-02 are strongest for $m_\pi$ above few GeV, just as they were
for the $SU(2)_d$ model in \cref{fig:constraints}. The constraints on the
$e^+e^-$ final state by Essig et al.~\cite{Essig:2013goa}
also provide interesting limits on the dark pion model,
but they are weaker than bounds from the CMB, which give the strongest
constraints at $m_\pi \lesssim \text{GeV}$.
The reason why CMB bounds are so powerful in the dark pion model, while being
subdominant in the $SU(2)_d$ model from \cref{sec:darkSU2} is of course
the different sign of $\Delta$: for $\Delta > 0$, as in the dark pion model,
DM annihilation can be significant even at very small DM velocity.

\Cref{fig:constraintpion} shows that the dark pion model is not
constrained by indirect searches for $m_\pi \gtrsim \mathcal{O}(1) \text{GeV}$
and $g' \in \left[10^{-3}, 1 \right]$, assuming annihilation to $\pi_d^0 \pi_d^0$
dominates.
The available parameter space is thus larger than for conventional WIMP
dark matter, which CMB constraints~\cite{Ade:2015xua} and Fermi dwarf galaxy
observations~\cite{Ackermann:2015zua} force to be heavier
than $\mathcal{O}(10 \sim 100)$\,GeV. This is the main success of the
Impeded Dark Matter paradigm.

\section{Conclusions}
\label{sec:conclusions}

\begin{table}
  \centering
  \footnotesize
  \centering
  \begin{tabular}{M{0.21\textwidth}|M{0.21\textwidth}|M{0.21\textwidth}|M{0.3\textwidth}}
  \hline \hline
    Model  & \multicolumn{2}{c|}{$SU(2)_d$ dark gauge boson} &  dark pion \\
    \hline
    \multirow{3}{*}{mass splitting}
      & \multicolumn{2}{c|}{$\Delta \simeq -\frac{1}{2} \eps^2 m_\text{DM}$,
        \quad \cref{eq:su2-delta}}
      & $\Delta \simeq g'^2 f^2_\pi / (2 m_\pi)$,
        \quad \cref{eq:pisplitting} \\
    \cline{2-4}
      & $10^{-7} \lesssim \eps \lesssim 10^{-3}$  & $\eps \gtrsim 10^{-3}$
        & $g' \gtrsim 0.05 $ \\
      & $\Delta < 0$  small & $\Delta < 0$ large
        & $\Delta > 0$ \\
    \hline
    freeze-out & \multicolumn{3}{c}{$\sigma v_\text{rel} \propto v_\text{rel}$} \\
    \hline
    CMB & $\sigma v_\text{rel} \simeq 0$ & \multirow{3}{*}{$\sigma v_\text{rel} \simeq 0$}
      & \multirow{2}{*}{$\sigma v_\text{rel} \propto \sqrt{\frac{2 \Delta}{m_\text{DM}}}$} \\
    \cline{1-2}
    Galaxies & \multirow{2}{*}{$\sigma v_\text{rel} \propto v_\text{rel}$} &  &  \\
    \cline{1-1} \cline{4-4}
    Clusters &  &   & $\sigma v_\text{rel} \propto \textbf{BF} \times
                       \sqrt{\frac{2\Delta}{m_\text{DM}}}$ \\
    \hline \hline
  \end{tabular}

  \caption{Mass splittings $\Delta$ and annihilation cross-sections $\sigma v_\text{rel}$
    for the two Impeded DM models discussed in this paper.
    In the $SU(2)_d$ dark gauge boson model, $\Delta$ depends on the
    kinetic mixing parameter $\eps$, while in the dark pion model it depends on the
    $U(1)'$ (dark electromagnetic) gauge coupling $g'$.
    Note that the annihilation cross-section in galaxies clusters receives a boost
    factor $(\textbf{BF})$ from halo substructure in the dark pion model, while a similar
    boost is absent in the $SU(2)_d$ model as $\sigma v_\text{rel}$ drops at small
    $v_\text{rel}$.}
  \label{tab:2models}
\end{table}

In summary, we have studied a class of dark matter models dubbed ``Impeded
DM'', which are characterized by a very small mass splitting $\Delta$ between
the DM and its annihilation products. $\Delta$ can be either
positive or negative.  For negative $\Delta$, Impeded DM is characterized by an
annihilation cross-section $\sigma v_\text{rel}$ that grows \emph{linearly}
with $v_\text{rel}$. This behavior allows for a regular thermal freeze-out,
while constraints from low-$v_\text{rel}$ environments (CMB, dwarf galaxies)
are suppressed.  For positive $\Delta$, the annihilation cross-section can be
suppressed by the small ratio $\Delta/m_{\text{DM}}$.

We presented two specific models that realize the Impeded DM phenomenology (see
\cref{tab:2models} for a summary of $\sigma v_{\text{rel}}$ for the two models
under different conditions).  In the first one, DM comes in the form of massive
gauge bosons associated with a dark sector $SU(2)_d$ group. When $SU(2)_d$ is
broken, the mass of one of the three gauge bosons is changed by a small amount,
typically upwards ($\Delta < 0$).  The lighter gauge bosons constitute the DM,
while the slightly heavier gauge boson interacts with the SM sector through a
non-Abelian kinetic mixing term induced by a dimension six operator.

In the second Impeded DM model, the dark matter is composite. It features a
confining gauge group $SU(N)$ and two species of dark quarks, which form dark
pions.  An additional $U(1)'$ (dark electromagnetism) splits the pion triplet
in such a way that the DM particles $\pi^\pm_d$ are typically heavier than the
neutral $\pi^0_d$, into which they annihilate. The dark and visible sectors are
coupled through kinetic mixing of the dark and visible photons.

For both models, we have presented detailed investigations of the phenomenology
and have constrained the parameter space using all available data from
cosmology, direct and indirect detection.

We conclude that Impeded DM populates a new niche in DM model space, but
a niche that is becoming more and more interesting as CMB and dwarf galaxy
constraints on DM annihilation put conventional thermal relic models
under severe pressure.

\section*{Acknowledgments}

JK would like to thank Fermilab and the Aspen Center for Physics
for hospitality and support during the final stages of this project,
as well as Lufthansa for discounted WiFi access and a power outlet
at a particularly critical time.
WX is grateful to the Mainz Institute for Theoretical Physics (MITP) for its hospitality
and its partial support during the completion of this work.
JL and XPW would like to thank Yang Bai for helpful discussion.
JL would like to thank Felix Yu for pointing out typos in Appendix B.
The work of JK, JL, and XPW is supported by the German Research
Foundation (DFG) under Grant Nos.\ \mbox{KO~4820/1--1} and FOR~2239 and by the
European Research Council (ERC) under the European Union's Horizon 2020
research and innovation programme (grant agreement No.\ 637506,
``$\nu$Directions''). The work of TS and WX is supported by the U.S. Department
of Energy under grant Contract Numbers DE$-$SC00012567 and DE$-$SC0013999.

\appendix

\section{DM annihilation to SM particles in the $SU(2)_d$ model}
\label{sec:appendixSM}

We list here the annihilation cross-sections for the processes
$K_1 K_2 \to u\bar{u}, d\bar{d}, e^+e^-, \nu \bar{\nu}, W^+W^-$
in the $SU(2)_d$ model
\begin{align}
  \sigma v_\text{rel}(K_1 K_2 \to u \bar{u})
    &= \frac{e^2g_d^2v^2\eps^2\sqrt{m_k^2-m_u^2}}{3888\pi \cos^4\theta_w m_k^5\left(4m_k^4-5m_k^2m_Z^2+m_Z^4\right)^2}\times\left\{64\cos^4\theta_w m_Z^8\left(2m_k^2+m_u^2\right) \right. \nonumber \\
&-32 \cos^2\theta_w m_k^2m_Z^6 (28\cos^2\theta_w-5)(2m_k^2+m_u^2)+ 16m_k^8(68m_k^2+7m_u^2)  \nonumber \\
&-16m_k^6m_Z^2\left[4m_k^2(70\cos^2\theta_w-17)+7m_u^2(20\cos^2\theta_w-1)\right]  \nonumber \\
&+ m_k^4m_Z^4\left[ 4\left(3008\cos^4\theta_w -2200 \cos^2\theta_w +833\right)m_k^2 \right. \nonumber \\
& +\left. \left( 6016\cos^4\theta_w -4400\cos^2\theta_w +343\right)m_u^2\right] \left. \right\}
  \,,\\
  \sigma v_\text{rel}(K_1 K_2 \to d \bar{d})
    &= \frac{e^2g_d^2v^2\eps^2\sqrt{m_k^2-m_d^2}}{3888\pi \cos^4\theta_w m_k^5\left(4m_k^4-5m_k^2m_Z^2+m_Z^4\right)^2}\times \left\{\right. 16\cos^4\theta_w  m_Z^8(2m_k^2+m_d^2)  \nonumber \\
& -16\cos^2\theta_w m_k^2m_Z^6 \left(14\cos^2\theta_w-1\right)\left(2m_k^2+m_d^2\right)+ 16 m_k^8\left(20m_k^2-17m_d^2\right)  \nonumber \\
&-16m_k^6m_Z^2\left[4m_k^2(7\cos^2\theta_w-5)+m_d^2(14\cos^2\theta_w+17)\right]  \nonumber \\
&+m_k^4m_Z^4\left[4m_k^2\left(752\cos^4\theta_w -220\cos^2\theta_w + 245\right) \right. \nonumber \\
&+ m_d^2 \left. \left(1504\cos^4\theta_w -440\cos^2\theta_w -833\right)\right] \left.\right\}
  \,,\\
  \sigma v_\text{rel}(K_1 K_2 \to e^+e^-)
    &= \frac{e^2g_d^2v^2\eps^2\sqrt{m_k^2-m_e^2}}{1296\pi \cos^4\theta_w m_k^5\left(4m_k^4-5m_k^2m_Z^2+m_Z^4\right)^2}\times \left\{\right. 16\cos^4\theta_w  m_Z^8(2m_k^2+m_e^2)  \nonumber \\
& -16\cos^2\theta_w m_k^2m_Z^6 \left(14\cos^2\theta_w-3\right)\left(2m_k^2+m_e^2\right)+ 16 m_k^8\left(20m_k^2+7m_e^2\right)  \nonumber \\
&-16m_k^6m_Z^2\left[4m_k^2(21\cos^2\theta_w-5)+7 m_e^2(6\cos^2\theta_w-1)\right]  \nonumber \\
&+m_k^4m_Z^4\left[4m_k^2\left(752\cos^4\theta_w -660\cos^2\theta_w + 245\right) \right. \nonumber \\
&+ m_e^2 \left. \left(1504\cos^4\theta_w -1320\cos^2\theta_w +343\right)\right] \left.\right\}
  \,,\\
  \sigma v_\text{rel}(K_1 K_2 \to \nu \bar{\nu})
    &= \frac{e^2g_d^2v^2\eps^2\sqrt{m_k^2-m_\nu^2} \left(4m_k^2-m_\nu^2\right) }{1296\pi \cos^4\theta_w m_k\left(4m_k^4-5m_k^2m_Z^2+m_Z^4\right)^2}
     \left(16m_k^2+16m_k^2m_Z^2 + 49 m_Z^4\right)
  \,,\\
  \sigma v_\text{rel}(K_1 K_2 \to W^+W^-)
    &= \frac{e^2g_d^2v^2\eps^2}{162\pi \cos^4\theta_w m_k^5\left(4m_k^4-5m_k^2m_Z^2+m_Z^4\right)^2} \left(94m_k^4-14m_k^2m_Z^2+m_Z^4\right) \nonumber \\
&\times \left(m_k^2-\cos^2\theta_wm_Z^2\right)^{\frac{3}{2}}\left(3\cos^4\theta_wm_Z^4+20 \cos^2\theta_wm_k^2m_Z^2+4m_k^4\right) \,,
\end{align}
where $v=v_\text{rel}/2$. We see that all of these cross-sections are proportional to $v_\text{rel}^2$, i.e.\
they are $p$-wave suppressed. Proportionality to $\eps^2$ leads to an additional
suppression.

The $p$-wave nature of $K_1 K_2$ annihilation to SM particles can be understood
by considering that all of the above processes involve the coupling
$(K_1^\mu \partial_\mu K_2^\nu - K_2^\mu \partial_\mu K_1^\nu)$.
In the non-relativistic limit, only contributions involving derivatives with respect
to time could in principle be unsuppressed by $v_\text{rel}^2$. These
contributions have the form $m_k (\xi_1^0 \xi_2^i - \xi_2^0 \xi_1^i)$,
where $\xi^\mu$ are the polarization vectors of the DM particles, and $i=1,2,3$.
This is a $p$-wave state \cite{Kumar:2013iva}, therefore the overall annihilation
cross-section must be $p$-wave suppressed.

\section{$K_3$ decay to SM particles in the $SU(2)_d$ model}
\label{sec:appendixK3}

The partial decay widths of $K_3$ for decays to SM particles are obtained from
the non-Abelian kinetic mixing term \cref{eq:nonAkm}, after removing
the mixing and rotating to mass eigenstates according to \cref{eq:su2-shifts}.
We find
\begin{align}
\Gamma(K_3\to u \bar{u})
&= \frac{\eps^2e^2\sqrt{m_k^2-4m_u^2}}{288\pi \cos^2\theta_w m_k^2(m_k^2-m_Z^2)^2}
\left[ 17 m_k^6 + m_k^4(7 m_u^2 - 40 \cos^2\theta_w m_Z^2 ) \right. \nonumber \\
& \left. + 16 m_k^2 (2\cos^4\theta_w m_Z^4 -  5 \cos^2\theta_w m_u^2 m_Z^2  )
+ 64\cos^4\theta_w m_u^2m_Z^4 \right]
\,,\\[0.2cm]
\Gamma(K_3 \to d \bar{d})
&= \frac{\eps^2e^2\sqrt{m_k^2-4m_d^2}}{288\pi \cos^2\theta_w m_k^2(m_k^2-m_Z^2)^2}
\left[ 5 m_k^6 + m_k^4(- 17 m_d^2 - 4 \cos^2\theta_w m_Z^2 ) \right. \nonumber \\
& \left. + 8 m_k^2 (\cos^4\theta_w m_Z^4 -   \cos^2\theta_w m_d^2 m_Z^2  )
+ 16 \cos^4\theta_w m_d^2m_Z^4 \right]
\,,\\[0.2cm]
\Gamma(K_3 \to e^+ e^-)
&= \frac{\eps^2e^2\sqrt{m_k^2-4m_e^2}}{96\pi \cos^2\theta_w m_k^2(m_k^2-m_Z^2)^2}
\left[ 5 m_k^6 + m_k^4(7 m_e^2 - 12 \cos^2\theta_w m_Z^2 ) \right. \nonumber \\
& \left. + 8 m_k^2 (\cos^4\theta_w m_Z^4 -  3 \cos^2\theta_w m_e^2 m_Z^2  )
+ 16 \cos^4\theta_w m_e^2m_Z^4 \right]
\,, \\[0.2cm]
\Gamma(K_3\to \nu \bar{\nu})
&= \frac{\eps^2e^2m_k^5}{96\pi \cos^4\theta_w (m_k^2-m_Z^2)^2}
\,, \\[0.2cm]
\Gamma(K_3\to W^+ W^-)
&=\frac{\eps^2e^2\sqrt{m_k^2-4\cos^2\theta_w m_Z^2}}{192\pi \cos^4\theta_w m_k^2(m_k^2-m_Z^2)^2} \nonumber \\
& \times \left(m_k^6+16\cos^2\theta_w m_k^4m_Z^2-68\cos^4\theta_w m_k^2m_Z^4-48\cos^6\theta_w m_Z^6\right) \,.
\end{align}

\bibliography{ref}
\bibliographystyle{utphys}

\end{document}